\documentclass[conference]{IEEEtran}
\IEEEoverridecommandlockouts

\usepackage[english]{babel} \usepackage[T1]{fontenc} \usepackage[utf8]{inputenc} \usepackage{graphicx}
\usepackage[hyphens]{url}
\usepackage{amsmath,amsfonts,amsthm}
 \usepackage{amssymb}
\usepackage[font=footnotesize,hangindent=10pt]{subfig}
\usepackage[linesnumbered,vlined,boxed,ruled]{algorithm2e}
\usepackage[table]{xcolor}
\usepackage{xparse}

\usepackage{wasysym}
\usepackage{tabu}
\usepackage{stix}

\usepackage{makecell}

\usepackage{caption}
\usepackage{pifont}
\usepackage{multirow}
\usepackage{enumitem}
\usepackage{booktabs}
\usepackage{threeparttable}
\usepackage{soul}
\usepackage[colorlinks,linkcolor=black,anchorcolor=black,citecolor=black]{hyperref}
\usepackage[capitalise]{cleveref}

\usepackage{marvosym}

\graphicspath{{./}{figure/}}
\crefformat{section}{\S#2#1#3}

\SetVlineSkip{2pt}

\SetCommentSty{myAlgComment}
\SetKwProg{Fn}{Function}{:}{}

\SetAlFnt{\small}
\SetAlCapFnt{\small}
\SetAlCapNameFnt{\small}
\crefname{algorithm}{Alg.}{Algs.}

\newcommand{\bullethdr}[1]{\noindent\textbullet\,\textbf{#1}}

\newcommand{\header}[1]{\smallskip\noindent\textbf{#1}}

\newcommand{\Real}{\mathbb{R}}

\newcommand{\CS}{\mathcal{S}}
\newcommand{\CB}{\mathcal{B}}

\newcommand{\CD}{\mathcal{D}}
\newcommand{\bv}{\mathbf{v}}
\newcommand{\bb}{\mathbf{b}}
\newcommand{\bh}{\mathbf{h}}
\newcommand{\bp}{\mathbf{p}}
\newcommand{\bW}{\mathbf{W}}

\IEEEoverridecommandlockouts
\makeatletter\def\@IEEEpubidpullup{3.5\baselineskip}\makeatother
\IEEEpubid{\parbox{\columnwidth}{
		Network and Distributed System Security (NDSS) Symposium 2027\\
		22-26 March 2027, Seoul, Republic of Korea\\ 
		ISBN 978-1-970672-09-1\\  
		https://dx.doi.org/10.14722/ndss.2027.233572\\
		http://www.ndss-symposium.org
}
\hspace{\columnsep}\makebox[\columnwidth]{}}

\title{MAppPrint: Robust and Real-time Mobile App\\
Fingerprinting via Traffic Behavior\\ Correlation Graphs
}

\author{  \IEEEauthorblockN{Tao Duan\IEEEauthorrefmark{2},\quad 
                    Junzhou Zhao\IEEEauthorrefmark{2}\textsuperscript{\scriptsize\Letter}\thanks{\textsuperscript{\scriptsize\Letter}Corresponding Author},\quad 
                    Jianglong Li\IEEEauthorrefmark{2},\quad 
                    Ronghua Li\IEEEauthorrefmark{3},\\
                    Youqian Zhang\IEEEauthorrefmark{3},\quad 
                    Nuo Xu\IEEEauthorrefmark{2},\quad 
                    Pinghui Wang\IEEEauthorrefmark{2}\textsuperscript{\scriptsize\Letter},\quad 
                    Haibo Hu\IEEEauthorrefmark{3}\IEEEauthorrefmark{4}
                    }
  
  \IEEEauthorblockA{\IEEEauthorrefmark{2}{MOE KLINNS Lab, Xi'an Jiaotong University},
  \IEEEauthorrefmark{3}{The Hong Kong Polytechnic University}\\
    \IEEEauthorrefmark{4}{PolyU Research Centre for Privacy and Security Technologies in Future Smart Systems}
}
 \IEEEauthorblockA{
                    \{duantao, ljl522426, xunuo\}@stu.xjtu.edu.cn, \{junzhou.zhao, phwang\}@xjtu.edu.cn,\\ cory-ronghua.li@connect.polyu.hk, \{you-qian.zhang, haibo.hu\}@polyu.edu.hk
\vspace{-1ex}
 }
}

\begin{document}

\maketitle

\begin{abstract}
Mobile apps have become ubiquitous in daily life.
Despite communication encryption,
app fingerprinting (AF) attacks can still enable adversaries to infer the apps in use by analyzing patterns of encrypted network traffic.
However, existing AF attacks suffer from three critical limitations that hinder their practical effectiveness: (i) the inability to capture app-specific traffic dependencies, rendering traffic fingerprints fragile under dynamic networking conditions; (ii) the neglect of latency and throughput considerations, resulting in inefficient real-time user tracking; and (iii) the incapacity to handle mixed traffic generated by multi-app concurrency.

In this paper, we propose MAppPrint, a robust and real-time app fingerprinting framework. MAppPrint introduces a behavior-level traffic fingerprint, i.e., Traffic Behavior Correlation Graph (TBCG), which models app traffic by capturing unit traffic behaviors and their correlations, thereby providing reliable app fingerprints. To enable real-time fingerprinting, MAppPrint incorporates a fingerprint compression algorithm that transforms each TBCG into a compact graph by selectively aggregating redundant information. Furthermore, MAppPrint employs an unsupervised decomposition strategy to disentangle mixed traffic generated by multi-app concurrency, enabling unified identification of both single-app and multi-app traffic.
Extensive experimental results demonstrate that MAppPrint achieves
$4.9 \sim 13.9\%$ improvement in AF precision and $3 \sim 6$ times improvement in timeliness over state-of-the-art methods.
Meanwhile, MAppPrint is robust in dynamic networking conditions, while achieving a throughput of $10.5$ GB/s and a prediction latency of $0.68$ seconds.

\end{abstract}

\section{Introduction}
\label{sec:introduction}

Mobile apps have undergone an explosive increase in recent years due to the
pervasiveness of mobile devices such as smartphones and tablets~\cite{apps2025}.
Despite the widespread use of encrypted communication, existing works
have demonstrated that mobile apps are still
vulnerable to network-based app fingerprinting (AF) attacks~\cite{van2020flowprint}.
In such attacks, adversaries infer apps~\cite{pham2021mappgraph,wang2020appnet} and even in-app activities~\cite{bozorgi2022still,conti2015analyzing} by analyzing encrypted traffic side-channel information~\cite{gu2019traffic} such as packet length and packet
inter-arrival times using pre-trained machine learning (ML) or deep learning (DL) models. 
This technique poses a significant threat to user privacy and security. 
For instance, a privacy-intrusive eavesdropper can monitor a user’s real-time behavioral states via AF, whereas a malicious attacker can employ AF to identify sensitive user activities (e.g., transactions and logins) and then launch precisely targeted phishing or session hijacking attacks~\cite{du2021moves,wang2024off}.

When people are using mobile apps, most apps continuously or intermittently
interact with remote servers, generating diverse network traffic flows to
load various in-app services and functions.
Early {\em single-flow} based AF attacks learn discernible traffic patterns from individual
network flows~\cite{liu2019fsnet,taylor2016appscanner,taylor2017appscanner+}. 
However, due to the limited fingerprint information contained within a single network flow, 
the fingerprinting accuracy of such single-flow based methods is suboptimal in practice~\cite{xie2023rosetta}.
To address this weakness, recent studies have proposed {\em multi-flow} based AF attacks~\cite{van2020flowprint,jiang2022fgnet,pham2021mappgraph},
which aggregate concurrent flows generated by an app within a time window to obtain a richer traffic fingerprint. 
Despite improvements in accuracy over single-flow based methods, 
three fundamental limitations of existing multi-flow based methods still hinder their reliable and real-time deployment in realistic environments.

\bullethdr{Poor Robustness to Dynamic and Ambiguous Traffic}.
In real-world AF, adversaries need to deploy trained AF models to target networks.
Consequently, the {\em robustness} of AF methods is crucial.
However, existing methods construct multi-flow fingerprints by extracting per-flow features independently and then aggregating them at the flow level~\cite{pham2021mappgraph,qu2023ha,xu2025flowminer}.
This paradigm essentially captures only coarse-grained flow concurrency patterns, failing to reveal the inherent dependencies among fine-grained behaviors (e.g., the specific  UI loading logic within an app).
In practice, such multi-flow fingerprints are susceptible to dynamic networking, as per-flow features are prone to drift caused by app updates, network flapping, or region changes. Meanwhile, these methods struggle to distinguish ambiguous traffic generated by functionally similar apps, which usually exhibit analogous flow concurrency patterns. 

\bullethdr{Conflict between Accuracy and Timeliness}.
  In daily life, users usually interact with apps in real time and may change their activities rapidly.
  Consequently, the {\em timeliness} of AF methods, including latency and throughput,
  is crucial for enabling real-time user state tracking and launching timely follow-up attacks (e.g., hijacking payment-related sessions).
  However, existing methods require long observation windows to capture a sufficient number of concurrent flows, leading to substantial identification latency~\cite{van2020flowprint}.
  Meanwhile, most existing methods assume that adversaries possess
  sufficient computational and storage resources, overlooking the actual
  throughput burden imposed by high-volume app traffic, rendering existing methods  inefficient for online fingerprinting.

\bullethdr{Blind Spot in Multi-app Identification}.
  Moreover, existing methods assume that users activate
    only one app at a time~\cite{van2020flowprint}.
  However, in reality, modern mobile devices widely support {\em multi-app}
  running mode, e.g., multi-window playing~\cite{AndroidMultiWindow} and
  background processing, allowing users to run multiple apps simultaneously on a single device.
  In multi-app running mode, each traffic trace is a mixture of multiple apps' traffic.
  Consequently, directly aggregating concurrent flows disrupts the fingerprint patterns of individual apps in such mixed traces, rendering existing methods ineffective in multi-app scenarios.

To address these limitations, we propose MAppPrint, a \emph{robust} and {\em
  real-time} \textbf{M}obile \textbf{App} finger\textbf{Print}ing attack.
  The core innovation of MAppPrint is a novel traffic fingerprint, called the Traffic Behavior Correlation Graph (TBCG), 
which models app traffic by extracting fine-grained unit traffic behaviors and capturing behavior dependencies using temporal correlations.
Our design is motivated by two key observations.
First, unit traffic behaviors, each composed of consecutive traffic segments in network flows, are the minimum units to carry
complete networking activities (e.g., messaging and UI loading), and usually
exhibit {\em stable} feature patterns.
Second, when users interact with an app, the correlations (e.g.,
response and trigger) between the generated unit traffic behaviors are closely
related to unique in-app operational logic and module dependencies, and thus can provide
{\em expressive} app fingerprints.
Different from existing multi-flow AF methods that rely on coarse-grained flow concurrency patterns, TBCG profiles the inherent traffic dependencies of an app at the behavior level.
It thus provides more reliable app fingerprints in dynamic and ambiguous traffic while reducing the reliance on lengthy traffic observation windows.

However, achieving real-time AF remains challenging due to the substantial computational overhead imposed by high-volume network traffic.
A simple yet effective solution is to compress the TBCG to obtain a concise app
fingerprint. Nevertheless, balancing fingerprint compression against the resulting
information loss is non-trivial.
Inspired by our in-depth study of app traffic traces, which reveals numerous
similar traffic behaviors in the generated traces due to the repeated invocations
of certain modules during runtime, we further propose a TBCG
compression algorithm.
Our algorithm selectively aggregates {\em redundant} nodes and edges in each TBCG based on behavior similarity so as to preserve critical fingerprint information, thereby improving system throughput with minimal impact on AF accuracy.

Finally, we propose an app identification framework to predict app labels.
A primary challenge in this stage is that the number of concurrent apps and their
respective traffic boundaries are unknown. 
To address this, we introduce an unsupervised decomposition strategy
that leverages inter-flow similarities to disentangle multi-app TBCGs
into individual single-app TBCGs. 
We then employ a lightweight model to provide a {\em unified} prediction of both single- and multi-app labels. 
By leveraging this framework, MAppPrint effectively fingerprints apps across diverse
running modes without requiring any prior knowledge.

\cref{tab:methods} compares MAppPrint with existing AF methods.
Our contributions can be summarized as follows:
\begin{itemize}[leftmargin=*, nosep]
\item We introduce MAppPrint, the first AF method that
  simultaneously reconciles accuracy, timeliness, and robustness.
  The key innovations of MAppPrint include a novel traffic fingerprint modeling
  approach (i.e., TBCG), a high-fidelity fingerprint compression algorithm,
  and a unified single-app and multi-app identification framework.

\item We develop an automatic app traffic triggering and labeling system, and collect
  and release the first complex-scenario AF benchmark.
  This benchmark contains over $450$ GB of app traffic traces, encompassing six
  real-world operating environments, including app updates, bandwidth/region changes, network flapping, and multi-app concurrency.

\item We conduct extensive experiments on $12$ datasets.
  The experimental results demonstrate that, compared to state-of-the-art
  methods, MAppPrint
  \footnote{Code and datasets are available at
    \url{https://github.com/ToDuan/MAppPrint}} 
  achieves a $4.9\sim
  13.9\%$ improvement in AF precision and a $3 \sim 6$ times improvement in timeliness.
  In addition, MAppPrint exhibits superior robustness in dynamic network environments, while achieving an
  average throughput of $10.5$ GB/s with a prediction latency of $0.68$ seconds in online deployment. 
\end{itemize}

\begin{table}[t]

  \centering
    \caption{Comparison of existing AF methods.
    Here \CIRCLE, \LEFTcircle, and \Circle{} denote the high, medium, and low
    performance levels.\label{tab:methods}}
  \begin{tabular}{c|@{}c@{  }@{}c@{}|@{}c@{}@{ }c@{}|@{}c@{  }}
    \bottomrule
    \multirow{3}{*}{AF methods} & \multicolumn{2}{c|}{Accuracy} & \multicolumn{2}{c|}{Timeliness}  & \multirow{3}{*}{Robustness} \\
    \cline{2-5}
        & \makecell{Single\\app} & \makecell{Multi-\\app}
    & \makecell{Low\\latency} & \makecell{High\\throughput} &  \\
    \hline
        AppScanner~\cite{taylor2016appscanner}
    & \LEFTcircle & \LEFTcircle & \Circle & \Circle & \Circle  \\
    AWF~\cite{awf2018}
    & \LEFTcircle & \LEFTcircle & \Circle & \Circle & \Circle  \\
    DappGraph~\cite{shen2021dapp}
    & \LEFTcircle & \LEFTcircle & \Circle & \Circle & \Circle  \\
    FS-Net~\cite{liu2019fsnet}
    & \LEFTcircle & \LEFTcircle & \Circle & \Circle & \Circle  \\
        FlowPrint~\cite{van2020flowprint}
    & \LEFTcircle  & \Circle  & \Circle & \Circle & \LEFTcircle   \\
    MappGraph~\cite{pham2021mappgraph}
    & \LEFTcircle  & \Circle  & \Circle & \Circle & \LEFTcircle   \\
    FG-Net~\cite{jiang2022fgnet}
    & \LEFTcircle  & \Circle  & \Circle & \Circle & \LEFTcircle  \\
    HA~\cite{qu2023ha} & \LEFTcircle & \Circle & \Circle   & \Circle & \LEFTcircle  \\
    STC-WF~\cite{tan2024inter} & \LEFTcircle & \Circle & \Circle   & \Circle & \LEFTcircle  \\
    FlowMiner~\cite{xu2025flowminer} & \LEFTcircle & \Circle & \Circle   & \Circle & \LEFTcircle  \\
    \cline{1-6}
    \textbf{MAppPrint} & \CIRCLE & \CIRCLE  & \CIRCLE   & \CIRCLE    & \CIRCLE  \\
    \toprule
  \end{tabular}
  \vspace{-2ex}
\end{table}

\section{Threat Model and Method Overview}
\label{threat_model}

In this section, we present our threat model and provide a
brief overview of MAppPrint.

\subsection{Threat Model and Design Goals}

\begin{figure}[t]
  \centering
  \includegraphics[width=0.9\linewidth]{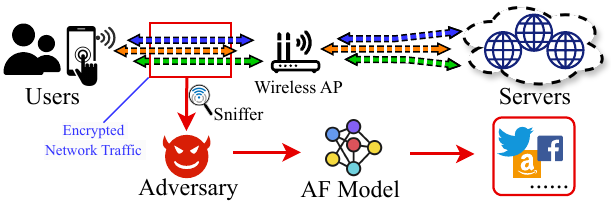}
  \caption{Threat model}
  \label{fig:threat_model}
  \vspace{-3ex}
\end{figure}

As shown in~\cref{fig:threat_model}, 
the adversary considered in this paper sniffs encrypted network traffic at the mobile network edge, such as wireless access points, switches, and gateways. 
The adversary's {\em goal} is to identify the apps in use from encrypted traffic, so as to deduce the target user's behavioral states or facilitate subsequent attacks. 
Consistent with prior AF methods~\cite{van2020flowprint,pham2021mappgraph,li2022packet}, we assume that the adversary cannot exploit packet payload or destination-related metadata (e.g., domain names and SNIs), as such information is usually encrypted. 
Before launching the attack, the adversary collects traffic samples of target apps in an offline environment to train the AF model. 
During the attack, the adversary deploys the trained model on the target network to analyze traffic traces.

\begin{figure*}[t]
  \centering
  \includegraphics[width=\linewidth]{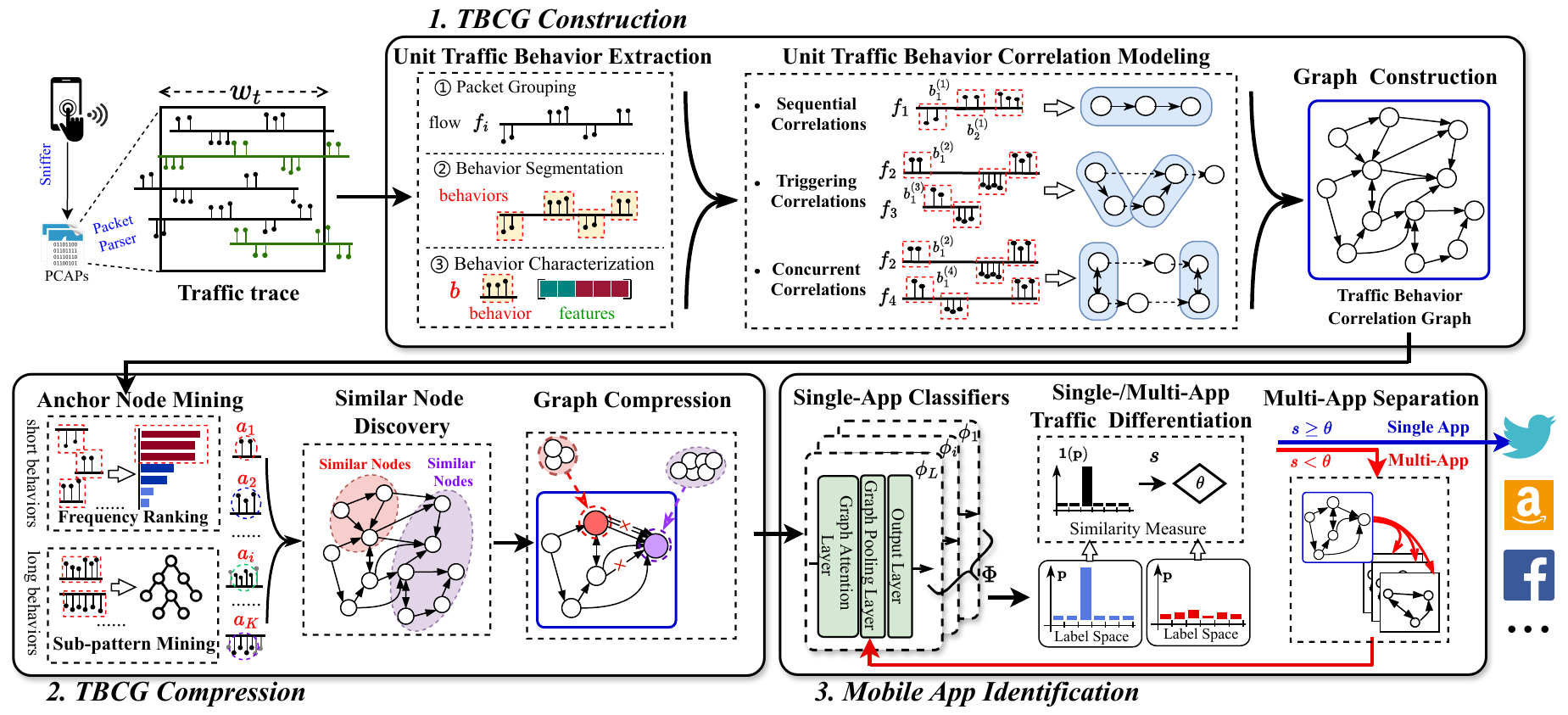}
  \caption{Overview of MAppPrint}
  \label{fig:overview}
 \vspace{-1ex}
\end{figure*}

Compared to existing methods, our threat model further incorporates the following real-world deployment conditions to establish more realistic AF attacks. 
First, we consider the {\em dynamics of network environments}, including app updates, region/network changes, and network flapping. 
This requires the AF system to model app traffic fingerprints that remain reliable across environments, thereby maintaining robust performance.
Second, we consider {\em real-time user tracking}, which requires the AF system to analyze high-speed and vast-volume app traffic online to respond to rapidly changing user behaviors and provide timely privacy information for subsequent attacks.
Third, we consider the {\em concurrent use of multiple apps}, where a user may simultaneously run multiple apps on a single device. 
Therefore, the traffic traces sniffed by the adversary may consist of either pure single-app traffic or mixed traffic generated by multiple apps. 
We emphasize that in our threat model, the adversary has no prior knowledge of the number of apps used by the target user.
This requires the AF system to uniformly handle both single- and multi-app traffic.

\subsection{Overview of MAppPrint}
\label{overview}

These real-world complexities present significant challenges to realistic AF.
Dynamic networking environments disturb the traffic patterns, gradually rendering previously learned app fingerprints obsolete.
The high-speed and vast-volume traffic traces incur substantial computational and storage overheads.
Concurrent apps directly disrupt the distinguishable fingerprints in traffic
traces.
To address these challenges, we develop MAppPrint for robust and real-time app fingerprinting.

\Cref{fig:overview} presents an overview of MAppPrint, which 
comprises three core modules, i.e., TBCG construction, TBCG
compression, and mobile app identification.
Given a traffic trace 
comprising multiple concurrent network flows, 
the TBCG construction module first extracts unit traffic behaviors and their correlations inherent to app operational logic, 
to model a reliable app traffic fingerprint, referred to as a TBCG. 
Subsequently, the TBCG compression module compresses each TBCG to a compact graph by aggregating
redundant nodes and edges, retaining only key structures to maximize the
system's online throughput.
Finally, the mobile app identification module employs a unified framework to
simultaneously handle single- and multi-app traffic traces.
Single-app TBCGs are directly identified, while multi-app TBCGs are decomposed
into several single-app TBCGs and then identified.

Note that unlike existing multi-flow AF attacks~\cite{jiang2022fgnet,xu2025flowminer}, which construct app fingerprints by aggregating traffic flows for offline single-app fingerprinting, 
MAppPrint extracts traffic behaviors as basic analysis units and captures behavior-level traffic dependencies, providing stable fingerprints for dynamic and ambiguous traffic.
Moreover, MAppPrint enables real-time fingerprinting of both single- and multi-app traffic through dedicated fingerprint compression and decomposition.

\section{Traffic Behavior Correlation Graph}
\label{sec:tbcg_construct}

In this section, we introduce the Traffic Behavior Correlation Graph (TBCG), which
can model behavior-level dependencies of app traffic.
To build a TBCG, we first extract the unit traffic behaviors from traffic
traces, which are the nodes in a TBCG.
Then, we define three fundamental correlations among unit traffic behaviors,
which form the edges in a TBCG.
Finally, to improve the computational efficiency over a TBCG, we design methods
to reduce the sheer size of a TBCG by merging redundant nodes and edges.

\subsection{Unit Traffic Behavior Extraction}
\label{ss:unit_behavior_extraction}

To model the network traffic generated by mobile apps, it is common to choose
either packets~\cite{qu2023ha} or flows~\cite{xu2025flowminer} as the basic
analysis units to extract per-packet or per-flow features.
However, as previously discussed, per-packet and per-flow features are susceptible to feature drift induced by dynamic networking environments.
Instead, we propose to use the {\em unit traffic behaviors} as the basic
analysis units.

Specifically, when people are interacting with a mobile app, rich traffic
behaviors are generated, accompanied by user activities
(e.g., sending messages, loading UI).
We observe that a traffic behavior is usually associated with a sequence of a few
consecutive packets with small inter-arrival times in a flow.
This type of packet sequence exhibits relatively stable patterns, and
has an explainable semantic meaning as well.
Hence, we choose such a kind of semantically complete packet sequence within a
traffic trace as the basic analysis unit when fingerprinting an app, and refer
to it as a {\em unit traffic behavior}.

Formally, let us consider a traffic trace generated by a mobile device in an
observation time window.
A traffic trace is represented as a sequence of packets $\CS\triangleq
\{pkt_1,pkt_2,\ldots\}$.
Each packet is denoted by $pkt\triangleq (key,len,dir,time)$, where $key$
denotes the five-tuple of the packet, i.e., source/destination IP, source/destination port, 
and transport layer protocol; $len$ denotes the packet
length in bytes; $dir$ denotes the packet direction, e.g., from client to server
({\em uplink}) or from server to client ({\em downlink}); $time$ denotes the
timestamp of the packet.
Packets sharing the same $key$ form a flow, and a bidirectional flow contains
packets of both uplink and downlink.
Let us denote the set of bidirectional flows in $\CS$ by $\{f_1, f_2, \ldots\}$.

A unit traffic behavior is defined as a sequence of a few consecutive packets with
small inter-arrival times that share the same direction in a bidirectional flow.
In the following discussions, a flow always means a bidirectional flow.
We denote the set of unit traffic behaviors in flow $f_i$ by $\CB_i \triangleq
\{b_1^{(i)}, b_2^{(i)},\ldots\}$, where $b_j^{(i)}\triangleq \{pkt_k^{(i)},
pkt_{k+1}^{(i)}, \ldots, pkt_s^{(i)}\}$ represents the $j$-th unit traffic
behavior in flow $f_i$, consisting of consecutive packets from the $k$-th to the
$s$-th packet in $f_i$. Each behavior is a compact burst of packets in a flow.
We give the detailed pseudo-code of extracting unit traffic behaviors in a flow
in \cref{alg:behavior_extraction} (see Appendix \cref{sec:implementation}).
Finally, the set of unit traffic behaviors in a traffic trace $\CS$ is denoted
by $\CB\triangleq\cup_i\CB_i$.

Notice that even though two unit traffic behaviors have the same semantic
meaning, e.g., text messaging, due to the difference of message content, these
two unit traffic behaviors may still not be exactly the same.
In order to facilitate subsequent analysis, we propose to transform each unit
traffic behavior $b\in\CB$ into a $12$-dimensional feature vector.
The feature vector consists of six transmission related features, three length
related features, and three time related features (see
\cref{tab:behavior_features} in Appendix \cref{sec:implementation}).
These features neglect specific packet details and emphasize the local
patterns of each unit traffic behavior from multiple perspectives
to mitigate the impact of dynamic network environments on the stability
of traffic fingerprints.

\subsection{Correlations of Unit Traffic Behaviors}
\label{ss:correlations_of_behaviors}

Till now a straightforward way to fingerprint an app is that we concatenate the
feature vectors of all unit traffic behaviors of an app to obtain a long
feature vector to represent the app.
However, if two apps share some common software modules, e.g., advertising
module and CDN module, similar unit traffic behaviors will be generated by both
apps, and this makes the two apps indistinguishable.

To address this issue, we consider to further explore the correlations of unit
traffic behaviors in an app.
Intuitively, even though different apps share some common unit traffic
behaviors, they may still have distinct {\em behavior correlations}.
This idea is motivated by the finding that different apps usually have
inherently different operational logic and software module dependencies determined by the UI ViewTrees~\cite{liu2019viewtree,zhang2017UIgraph,dong2018viewtree}.

For two unit traffic behaviors $b_r^{(i)}, b_{s}^{(j)}\in\CB$ extracted from a
traffic trace, we define three fundamental correlations between them, namely,
the sequential correlation, the triggering correlation, and the concurrent
correlation.

\begin{itemize}[leftmargin=*, nosep]
\item \textbf{Sequential Correlation}: If $b_r^{(i)}$ and $b_{s}^{(j)}$ belong
  to the same flow and they are temporally consecutive with inter-arrival time less
  than a threshold, i.e., $i=j$ and $s=r+1$ hold, then we say that a sequential
  correlation exists between the two unit traffic behaviors, denoted by
  $b_r^{(i)}\rightarrow b_{s}^{(j)}$.
  For example, in \cref{fig:overview}, we have $b_1^{(1)}\rightarrow b_2^{(1)}$.
\item \textbf{Triggering Correlation}: If $b_r^{(i)}$ and $b_{s}^{(j)}$ belong
  to two different flows, i.e., $i\neq j$, they are temporally consecutive with no
  overlap of lasting times, and their inter-arrival time is less than a
  threshold, then we say that a triggering correlation exists between
  $b_r^{(i)}$ and $b_{s}^{(j)}$, denoted by $b_{r}^{(i)} \rightsquigarrow
  b_{s}^{(j)}$.
  For example, in \cref{fig:overview}, we have $b_1^{(2)}\rightsquigarrow
  b_1^{(3)}$.
\item \textbf{Concurrent Correlation}: If $b_r^{(i)}$ and $b_{s}^{(j)}$ belong
  to two different flows, i.e., $i\neq j$, they are temporally consecutive, and have
  an overlap in lasting times, then we say that a concurrent correlation exists
  between the two unit traffic behaviors, denoted by $b_{r}^{(i)}
  \leftrightarrow b_{s}^{(j)}$.
  For example, in \cref{fig:overview}, we have $b_1^{(2)}\leftrightarrow b_1^{(4)}$.
\end{itemize}

The above three fundamental correlations imply the inner- and inter-flow
correlations of unit traffic behaviors generated by an app.
Specifically, sequential correlation represents the request-response
communications in a flow, which can capture the client-server interaction
patterns.
Triggering correlation represents the causal relationships between flows, which
can capture the functional dependencies between app modules.
For instance, an online payment request behavior will trigger an authentication
behavior, and the actual payment behavior only occurs after the successful
authentication.
Concurrent correlation represents the inter-flow collaboration patterns, which
can capture an app's response mechanism to simultaneous requests for multiple
resources.
For example, when refreshing a UI containing text, image, and video,
multiple requesting behaviors from different flows are generated simultaneously.

\subsection{TBCG Construction}

We can now build a Traffic Behavior Correlation Graph (TBCG), denoted by
$G=(V,E)$.
Here, graph $G$ is a directed acyclic graph; $V$ denotes the set of nodes,
representing the unit traffic behaviors extracted from a traffic trace; $E$
denotes the set of edges, representing the correlations among unit traffic
behaviors.

For each node $v\in V$, we represent it as a $12$-dimensional vector, denoted by
$\bv\in\Real^{12}$, as explained in \cref{ss:unit_behavior_extraction}.
An edge $(v_i, v_j)$ exists in the graph if $v_i$ and $v_j$ have any one of
the three correlations described in \cref{ss:correlations_of_behaviors}.
Notice that for a concurrent correlation between $v_i$ and $v_j$, edges $(v_i,
v_j)$ and $(v_j, v_i)$ both exist in the graph.
We give the detailed pseudo-code of constructing a TBCG in
\cref{alg:tbcg_construction} (see Appendix \cref{sec:implementation}).

\subsection{TBCG Compression}
\label{ss:tbcg_compress}

\begin{figure}[htp]
  \centering
  \vspace{-2ex}
  \subfloat[Nodes number distribution]{\includegraphics[width=.48\linewidth]{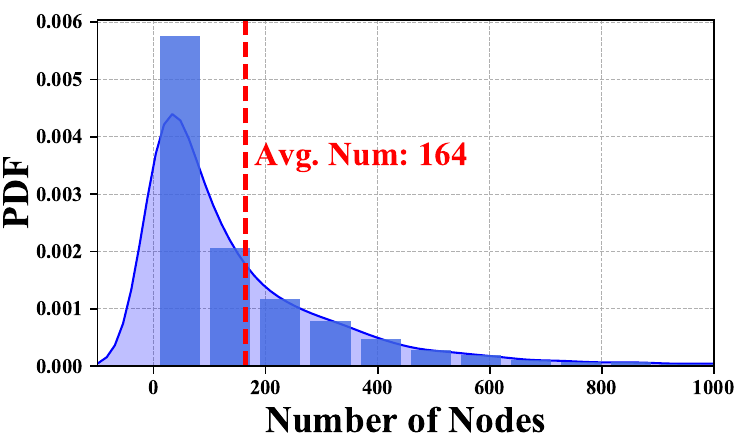}}
  \subfloat[Edges number distribution]{\includegraphics[width=.48\linewidth]{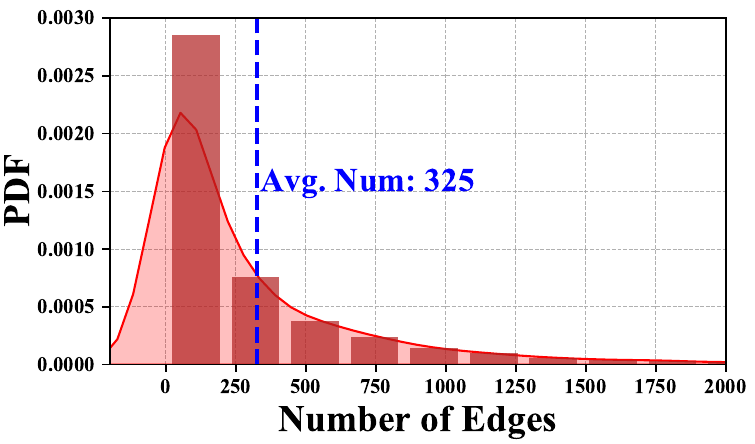}}
  \caption{Statistics on sizes of TBCGs}
  \label{fig:tbcg_sizes}
  \vspace{-1ex}
\end{figure}

So far we convert a traffic trace generated by a mobile device to a
directed graph, i.e., a TBCG, which captures traffic behaviors and their correlations.
If we directly leverage this TBCG to fingerprint an app, say, through learning a
graph neural network, we find that this approach is inefficient.
The reason is that the high-volume traffic app will result in a TBCG built even on a short traffic trace could contain a large number of nodes and edges.
For example, in \cref{fig:tbcg_sizes}, we collected $3,000$ ten-second duration
short traffic traces from Facebook, and find that the constructed TBCGs have
$164$ nodes and $325$ edges on average, and some graphs may even have thousands
of nodes and edges.
Learning on a large graph will harm the efficiency of app fingerprinting
significantly, and in practice, real-time app fingerprinting is important.
Therefore, it is necessary to find a way to reduce the sheer size of a TBCG
without losing too much accuracy in the successive analysis on the smaller
graph.

To this end, we present a graph compression algorithm to reduce the graph size.
Since compression inevitably causes information loss, a key consideration here is how to minimize the impact of TBCG compression on fingerprinting accuracy.
To motivate this algorithm, we analyze a real traffic trace generated by an
e-commerce app in \cref{tab:behaviors}.
We show six unit traffic behaviors belonging to two categories, i.e.,
TextSending and ProductBrowsing, and each is represented by a packet length
sequence.
We observe repeated sub-sequences in the packet length sequences belonging to
similar unit traffic behaviors due to repeated invocations of the same app modules at runtime.
For example, the TextSending category has a common sub-sequence ``$66\uparrow,
40\uparrow, 296\uparrow, 78\uparrow$'' and the ProductBrowsing category has a
common sub-sequence ``$1452\downarrow, 1278\downarrow, 478\downarrow$''.
Although these semantically similar behaviors may carry different content, such content-level differences are redundant for app identification. This motivates us to merge similar unit traffic behaviors by exploiting their common sub-sequences, thereby reducing redundancy while preserving critical fingerprint information.

\setlength{\fboxsep}{-0.5pt}

\begin{table}[htp]
  \centering
  \caption{Packet length sequences in unit traffic behaviors.
    Here ``$\uparrow$'' (``$\downarrow$'') denotes the uplink (downlink)
    packets.
    \label{tab:behaviors}}
  \begin{tabular}{c|c}
    \bottomrule
    Unit traffic behavior & Packet length sequence\tnote{1} \\
    \hline
    TextSending-1 & $634\uparrow$,\fcolorbox{red}{white}{$\mathbf{66}\uparrow,\mathbf{40}\uparrow,\mathbf{296}\uparrow,\mathbf{78}\uparrow$},$125\uparrow$      \\
    TextSending-2            & \fcolorbox{red}{white}{$\mathbf{66}\uparrow,\mathbf{40}\uparrow,\mathbf{296}\uparrow,\mathbf{78}\uparrow$},$133\uparrow,762\uparrow$    \\
    TextSending-3           & \fcolorbox{red}{white}{$\mathbf{66}\uparrow,\mathbf{40}\uparrow,\mathbf{296}\uparrow,\mathbf{78}\uparrow$},$998\uparrow$     \\ \hline
    ProductBrowsing-1         & $1452\downarrow$,\fcolorbox{blue}{white}{$\mathbf{1452}\downarrow,\mathbf{1278}\downarrow,\mathbf{478}\downarrow$},$296\downarrow$ \\
    ProductBrowsing-2         & $1271\downarrow$,\fcolorbox{blue}{white}{$\mathbf{1452}\downarrow,\mathbf{1278}\downarrow,\mathbf{478}\downarrow$},$447\downarrow$  \\
    ProductBrowsing-3         & $1278\downarrow$,\fcolorbox{blue}{white}{$\mathbf{1452}\downarrow,\mathbf{1278}\downarrow,\mathbf{478}\downarrow$},$484\downarrow$ \\
    \toprule
  \end{tabular}
\end{table}

To find similar nodes in a TBCG, a straightforward solution would be to compute
the pairwise packet-length sequence similarity of two nodes using a sequence
similarity computation algorithm such as the classic DTW
algorithm~\cite{eamonn2002dtw}.
However, calculating all pairwise node similarities has a computational
complexity of $O(|V|^2\alpha)$, where $\alpha$ is the time complexity of
calculating the similarity of two nodes, and this approach is too expensive for
online AF.

To compress a TBCG efficiently, we propose a fast node similarity computation
method.
The idea is that we choose a set of $K\ll |V|$ representative anchor nodes in
advance, and instead of computing all pairwise node similarities, we compare
each node with $K$ anchor nodes.
If two nodes are both similar to an anchor node, then these two nodes are
similar.
This approach reduces the computational complexity to $O(K|V|\alpha)$.
Meanwhile, the ratio $K/|V|$ determines the trade-off between compression efficiency and AF accuracy. Empirically, we find that setting $K/|V| \in [0.1, 0.2]$ can provide an effective balance.

\begin{figure}[t]
  \centering
  \subfloat[Length distribution]{\includegraphics[width=0.48\linewidth]{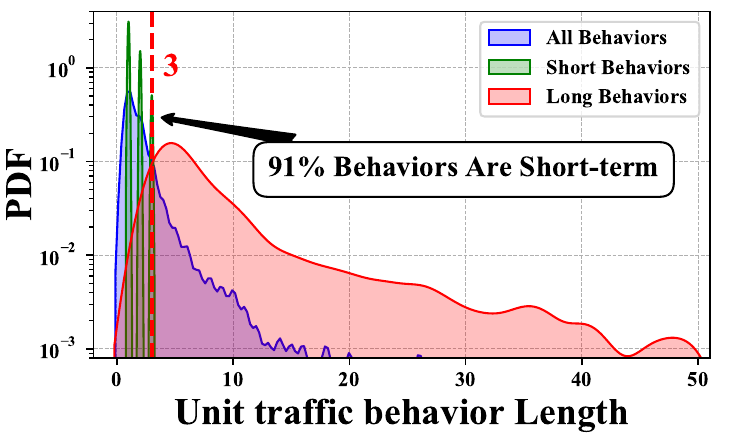}}
  \subfloat[Byte distribution]{\includegraphics[width=0.48\linewidth]{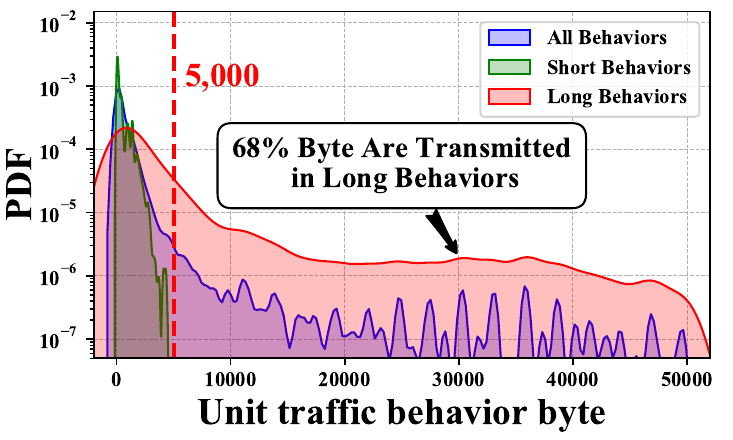}}
  \caption{Statistics of unit traffic behaviors}
    \label{fig:behaviors}
    \vspace{-2ex}
\end{figure}

\header{Anchor Node Discovery}.
To obtain the anchor nodes, we first show the distributions of the number of
packets and bytes in $6,000$ unit traffic behaviors collected from
Facebook in \cref{fig:behaviors}.
We find that a majority of behaviors contain no more than three
packets, which account for $91\%$ but just transmit $32\%$ bytes.
We refer to unit traffic behaviors with no more than three packets as short
behaviors, and the others as long behaviors.
The short behaviors are likely to be related to networking communications such
as handshaking and acknowledgment,
while the long behaviors are likely to be related to content transmissions such
as text or image downloading.

According to the difference of the number of packets in a unit traffic behavior,
we consider two types of anchor nodes discovery methods.
For short behaviors with no more than three packets, we directly rank them by
frequency and choose the top $0.9K$ most frequent unit traffic behaviors as
anchor nodes.
Due to the small number of packets a short behavior has, we do not further
choose sub-sequences in it but use the full packet length sequence for node
similarity computation.
While for long behaviors, we apply the PrefixSpan
algorithm~\cite{han2001prefixspan} to choose the top $0.1K$ most frequent packet
length sub-sequences as anchors.
Combining the results from both strategies, we can obtain a set of anchor nodes,
denoted by $A=\{a_1,\dots,a_K\}$.
Note that this process is completed during the offline initialization phase
before model deployment, thus incurring no online computational overhead.

\header{Similar Nodes Discovery and Graph Compression}.
Given the anchor node set $A$, each node in a TBCG is compared with these
anchors.
A node is considered to be similar to an anchor $a\in A$ if they are both short
or long behaviors and share the packet length sub-sequence in $a$.
Nodes in a TBCG that are similar to the same anchor are identified as similar
nodes and are merged into a single pseudo-node, as illustrated in
\cref{fig:overview}.
The feature vector of the merged pseudo-node is the mean of all node feature
vectors belonging to the same similar nodes group.
Meanwhile, each pseudo-node inherits all edges from its constituent nodes to
preserve the critical structure of the original graph.
Finally, duplicate edges are eliminated to further minimize redundancy.

\section{Mobile App Identification}
\label{sec:app_identification}

In this section, we identify apps from a traffic trace by exploiting the TBCG.
As discussed in \cref{sec:introduction}, in modern mobile devices, several apps
may run individually one at a time, or concurrently in the multi-window or background mode.
In the former case, the observed traffic trace in a short time window contains
only one app's traffic, while in the latter case, the observed traffic trace is
a mixture of multiple apps' traffic, which complicates the mobile app
identification task.

\subsection{Key Idea}
\label{ss:idea}

As a traffic trace may be generated by multiple apps, a straightforward solution
to identify apps from a traffic trace would be to treat the problem as a
multi-label classification problem, such as the approach taken by multi-tab
website fingerprinting~\cite{yin2021multi-tabWF,deng2023robustWF}.
This requires us to train a model that predicts a single label for a trace
containing a single app and multiple labels for a trace containing multiple
apps.
However, training this model relies on the availability of a large amount of
labeled traffic traces in both the single-app and multi-app scenarios.
While labeled single-app traffic traces can be easily collected, it is expensive
to collect and label multi-app traffic traces due to combination explosion.

If we can reduce a multi-app identification problem to multiple single-app
identification problems, then our problem will be greatly simplified.
To materialize this idea, we need to address the following two problems:
\begin{itemize}[nosep]
\item How can we determine whether an observed traffic trace is generated by a
  single app or multiple apps?
\item If a traffic trace is generated by multiple apps, how can we separate out
  the traffic of each app?
\end{itemize}

\begin{figure}[htp]
  \vspace{-2ex}
  \centering
  \subfloat[Single-app traffic]{\includegraphics[width=0.48\linewidth]{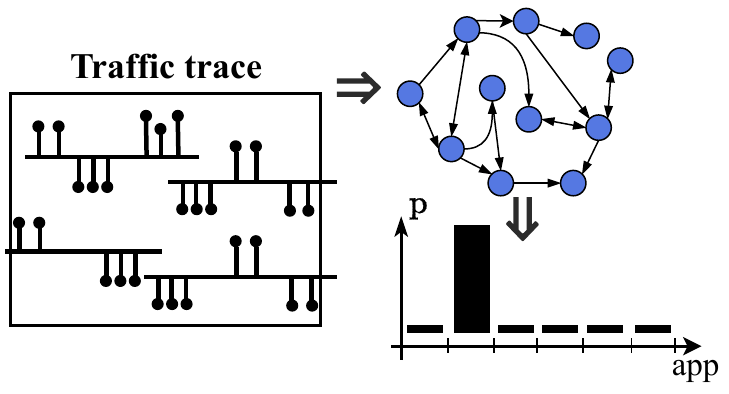}}
  \subfloat[Multi-app traffic]{\includegraphics[width=0.48\linewidth]{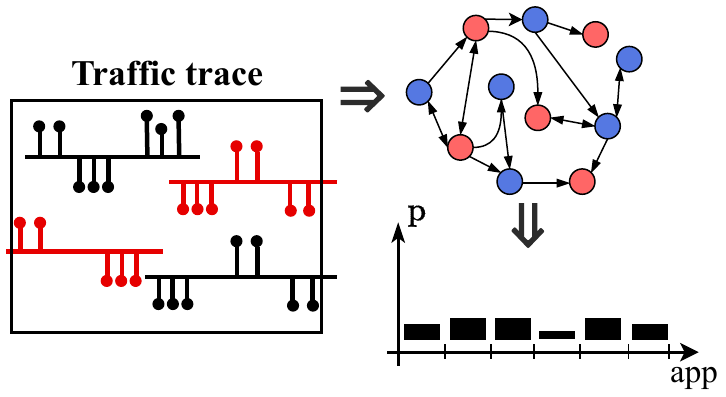}}
  \caption{Difference of single- and multi-app traffic\label{fig:app_mode}}
  \end{figure}

To address the first problem, we train multiple single-app traffic classifiers,
one for each app, and observe the difference of the probability distribution of
predicted label between single-app traffic and multi-app traffic.
As illustrated in \cref{fig:app_mode}, for single-app traffic, the probability
distribution of predicted label is unimodal, while for multi-app traffic, the
distribution is flattened.
This finding can help us to differentiate between single-app traffic and
multi-app traffic.
We elaborate our solution in \cref{ss:single_or_multi}.

To address the second problem, existing research has found that network traffic
generated by the same app typically exhibits similar flow-level transmission and
interaction patterns~\cite{prograph2023}.
For example, compared to a chat app, the flows generated by a video
app consistently exhibit lower interaction frequencies but larger
transmission payloads.
Thanks to the expressiveness of our proposed TBCG, which preserves the rich
interaction patterns of an app, we can exploit the similarity between sub-graph
structures within a TBCG to separate out each app's fingerprint.
We elaborate our solution in \cref{ss:reduction}.

Before elaborating on our solutions, we first propose a
method to identify the app from single-app traffic, which turns out to be the
basic building block to address the first problem.

\subsection{App Identification from Single-App Traffic}
\label{sec:classifier}

Given a TBCG $G$ constructed from a single-app traffic trace, we aim to obtain
the probability distribution of the predicted label of $G$.
We build a series of GNN-based {\em one-vs-rest} classifiers, one for each app,
denoted by $\Phi = \{\phi_1,\dots,\phi_L\}$, where $L$ denotes the number of
apps. Compared to traditional multi-class classifiers, one-vs-rest classifiers can
more reliably reject multi-app traffic that is not seen during training, rather
than misclassifying it as a single app~\cite{zhao2024trident}.

Each classifier $\phi_i$ consists of a graph attention layer, a graph
pooling layer, and an output layer, tailored to a specific app.

\header{Graph Attention Layer}.
A graph attention (GAT) layer~\cite{velickovic2018gat} is employed to adaptively
aggregate node information in the TBCG to learn app-specific node embeddings.
Specifically, for each node $v_i\in G$, the GAT layer quantifies the importance
of each neighbor node $v_j\in N_i$ to $v_i$ by $  \alpha_{ij} =
  \frac{\exp(\varphi(\bW_a[\bW\bv_i\Vert\bW\bv_j]))} {\sum_{v_k\in
      N_i}\exp(\varphi(\bW_a[\bW\bv_i\Vert\bW\bv_k]))}$,
where $\bW\in\Real^{d' \times d}$, $\bW_a\in\Real^{d' \times 2d'}$ are learnable
weight matrices, providing node feature linear transformations and attention
operation, respectively; $d$ and $d'$ are the dimensions of the input node
features and the corresponding latent vectors, respectively; $\varphi$ and
$\Vert$ denote the ReLU activation and vector concatenation operation,
respectively.
The representation of node $v_i$ is then refined through an aggregation of
neighborhood information, expressed as $\bv_i' = f(\sum_{v_j\in
  N_i}\alpha_{ij}\bW\bv_j)$, where $f$ is the activation function.

\header{Graph Pooling Layer.}
Next, the representation of TBCG $G$ is obtained through a
pooling operation $\bh_G = \frac{1}{|V|}\sum_{v_i\in G}\bv_i'$.

\header{Output Layer.}
Finally, the output layer of classifier $\phi_i$ calculates the acceptance
probability $p_i=\sigma(\bW_c\bh_G+\bb_c)$, denoting the probability that
classifier $\phi_i$ classifies $G$ as the $i$-th app.
Here, $\bW_c$ and $\bb_c$ are learnable parameters, and $\sigma$ denotes the
sigmoid activation function.

Each binary classifier is optimized independently on the corresponding training
set (see \cref{alg:training} in Appendix \cref{sec:implementation}).
The outputs from these $L$ binary classifiers are
stacked to form an $L$-dimensional vector.
The vector is further normalized to obtain a probability distribution $\bp$, and
slightly abusing the notations, we denote by $\bp = [p_1, \dots, p_L]^\top$,
where $p_i$ denotes the confidence score of $G$ belonging to the $i$-th app.

\subsection{Differentiating Single-/Multi-App Traffic}
\label{ss:single_or_multi}

We now elaborate on how to differentiate single-app and multi-app traffic to
address the first problem raised in \cref{ss:idea}.
As mentioned previously, for a single-app traffic trace, the probability
distribution of the predicted label $\bp$ is often unimodal, while for a multi-app
traffic trace, the distribution is flattened.
Inspired by this observation, we propose to compare the similarity of $\bp$ to a
one-point probability distribution, which exhibits an ideal unimodal
distribution.

Specifically, for each $\bp$, we first compute the corresponding one-point
probability distribution $\mathbf{1}(\bp)\in \Real^L$, which has a peak value of
$1$ at the dimension where $\bp$ achieves the maximum, and $0$ elsewhere.
The similarity between $\bp$ and $\mathbf{1}(\bp)$ is defined as $s\triangleq
1/(1+\chi^2(\bp, \mathbf{1}(\bp)))$, where $\chi^2(\cdot,\cdot)$ denotes the
$\chi^2$ distance.
A larger $s$ indicates a higher likelihood that $G$ corresponds to a single app,
and vice versa.
Let $\theta = \mu - 3\delta$ denote the similarity threshold, where $\mu$ and
$\delta$ represent the mean and standard deviation of $s$ obtained from all
labeled single-app traffic traces, respectively.

For a testing traffic trace, if it is identified as a single-app trace
(i.e., $s \geq \theta$), the MAppPrint workflow terminates, and its label is
predicted as $\hat{y} = \arg\max_i p_i$.
Conversely, if it is identified as a multi-app
trace (i.e., $s < \theta$), it will undergo further processing.

\subsection{Reducing Multi-App Traffic Identification to Single-App Traffic
  Identification}
\label{ss:reduction}

We next elaborate our approach to reduce the multi-app traffic identification
problem to multiple single-app traffic identification problems, and solve the
second problem raised in \cref{ss:idea}.
Note that the TBCG $G$ constructed from a multi-app traffic trace consists of
multiple apps' traffic patterns, and our goal is to separate out each app's
traffic pattern from $G$.

Specifically, we decompose $G$ into multiple subgraphs $\{g_1,g_2,\dots\}$ by
removing all triggering and concurrent correlations in $G$.
This step removes inter-flow correlations from $G$ so that the resulting
subgraph $g_i$ belongs to a single app and characterizes an app's flow-level
interaction and transmission patterns.
Each subgraph is further represented as a $10$-dimensional feature vector (see
\cref{tab:subgraph_features} in Appendix \cref{subsec:subgraph}).
Because different subgraphs may belong to the same app, we apply the
DBSCAN~\cite{schubert2017dbscan} algorithm on these subgraph feature vectors to
obtain several clusters.
Finally, all subgraphs belonging to the same cluster are merged and
reconstructed into a new single-app TBCG by inheriting the edges among the corresponding nodes
in the original $G$. 

After this step, a multi-app TBCG $G$ is divided into a set of single-app TBCGs,
denoted by $\{G_i'\}_{i=1,2,\ldots}$.
We then apply our single-app classifier on each $G_i'$ to
predict its label $\hat{y}_i$.

\section{Experiments}
\label{sec:experiments}

In this section, we evaluate the performance of MAppPrint and compare it with
state-of-the-art traffic fingerprinting methods to answer the following research
questions.

\begin{description}[nosep]
\item[RQ1] Can MAppPrint effectively fingerprint mobile apps in single-app scenarios? (\cref{subsec:single_app_evaluate})
\item[RQ2] Can MAppPrint effectively fingerprint mobile apps in multi-app scenarios? (\cref{subsec:multi_app_evaluate})
\item[RQ3] How robust is MAppPrint in dynamic networking environments, and
  why is it more robust? (\cref{subsec:robust_evaluate})
\item[RQ4] How efficient is MAppPrint for real-time app fingerprinting, and
  why is it more efficient? (\cref{subsec:efficiency_evaluate})
\end{description}

\subsection{Experimental Setup}

\header{Implementation.}
We prototype MAppPrint using Python 3.8 with more than $3500$ lines of code.
MAppPrint utilizes TShark for packet parsing and PyTorch
for model development.
We conduct five-fold cross-validation on each dataset and report the average
performance.
\cref{tab:gnn_architecture,tab:parameter} (see Appendix \cref{apdx:experiment_setup}) provide the detailed GNN architecture
 and default hyperparameter configurations, respectively.

\header{Baselines and Metrics.}
We select state-of-the-art traffic fingerprinting methods as baselines,
including $4$ single-flow methods, i.e.,
{\em AppScanner}~\cite{taylor2016appscanner}, {\em AWF}~\cite{awf2018},
{\em FS-Net}~\cite{liu2019fsnet}, and {\em DappGraph}~\cite{shen2021dapp}, and $4$
multi-flow methods, i.e., {\em FlowPrint}~\cite{van2020flowprint},
{\em MappGraph}~\cite{pham2021mappgraph}, {\em FG-Net}~\cite{jiang2022fgnet}, and
{\em FlowMiner}~\cite{xu2025flowminer}.
In the experiments, we use classic classification metrics, Precision and F1 Score,
to compare the performance of different methods, where higher values indicate better performance.

\header{Public Datasets.}
We first use three public mobile app traffic datasets to evaluate the performance of MAppPrint.
Our selection of datasets considers: (i) different app {\em scale}, (ii) different traffic {\em triggering methods} (i.e., triggered by app testing tools and real users), and (iii) varying degrees of {\em traffic ambiguity}. These datasets include:

\begin{itemize}[leftmargin=*, nosep]
    \item \underline{{\em FOAP}}~\cite{li2022foap}. This dataset collects traffic traces from $1,000$ apps on Google Pixel smartphones using the UI-touch tool~\cite{Monkey}. Due to varying collection durations, ranging from several minutes to several hours for different apps, we select the top-$120$ apps with the longest collection times to ensure sufficient traffic traces.
    
    \item \underline{{\em MAPP}}~\cite{pham2021mappgraph}. This dataset comprises traffic traces from the top-$101$ apps on the Google Play Store, collected by volunteers using Samsung, Xiaomi, and Google Pixel smartphones. All traffic traces are triggered by authentic user interactions.
   
    \item \underline{{\em SIM-APP}}~\cite{pham2021mappgraph}. This dataset collects traffic traces from $17$ functionally similar online audio apps using the same devices and traffic triggering method as the MAPP dataset.
    The traffic traces in this dataset exhibit significantly higher similarity than other datasets. 
    We use this dataset to evaluate MAppPrint's performance in ambiguous traffic.
\end{itemize}

\header{Newly Collected Datasets.}
Note that existing public datasets only collect traffic from single apps in static environments;
we collect $6$ new datasets to evaluate the performance of different methods in more realistic {\em dynamic networking} and {\em multi-app concurrent} scenarios. 
To achieve automatic large-scale traffic collection while maintaining consistency with actual user habits, we develop an app traffic triggering and labeling system (detailed in Appendix \cref{apdx:experiment_setup}).
Specifically, we first recruit volunteers to use mobile apps on smartphones
and record their activities. Subsequently, our system converts these real user activities into executable Android commands, combines different activities to generate continuous activity flows, and uses activity flow commands to control the UI-touch tool~\cite{Monkey} to trigger apps on different smartphones. 
Our collection includes:

\begin{itemize}[leftmargin=*, nosep]
    \item  Single-App Multi-Flow Dataset (\underline{{\em SAMF}}). We select the top-$100$ apps from the Google Play Store, remove e-banking and payment apps (which require user identity authentication), and use the remaining $94$ apps to collect single-app traffic. For each app, we design $7$ different interaction activities (e.g., scrolling the screen, sending text, watching videos, etc.), with each activity lasting about $30$ seconds and repeated $50$ times, and randomly combine these activities with one another to generate activity flows. In this dataset, each app is independently activated and triggers traffic through different activity flow commands.
    
    \item Multi-App Multi-Flow Dataset (\underline{{\em MAMF}}). 
    In this dataset, we combine the above apps running in the foreground and background of smartphones to collect multi-app traffic.
    Specifically, foreground apps are opened using multi-window (i.e., split-screen) mode,
    and are triggered by reusing activity flows from the SAMF dataset.
    Background apps are kept active using online modes 
        (e.g., video/audio playback, file upload/download)
    to prevent their processes from being killed. We note that, due to resource contention mechanisms, the actual number of apps that can run simultaneously in the background is limited. As illustrated by our analysis in~\cref{fig:traffic_pattern} (see Appendix \cref{sec:add_results}), in extreme scenarios ($5\sim 10$ concurrent apps), most background apps are idle, generating extremely sparse traffic. 
            Therefore, we set the number of concurrent apps to $2\sim 5$. 
       
    \item Dynamic Networking Environment Datasets. We collect app traffic in four representative dynamic networking environments (i.e., app updates, region or network changes, and network flapping) to evaluate the robustness of MAppPrint. The triggering method for apps remains consistent with that of the SAMF dataset.
    The \underline{{\em SAMF-Drift}} dataset collects traffic traces from the above $94$ apps
    continuously over a $120$-day period, during which most apps underwent at least
    one update.
    The \underline{{\em SAMF-Region}} dataset collects traffic traces of these $94$ apps from
    three different cities, separated by approximately $50$ to $1,000$ km.
    The \underline{{\em SAMF-Net}} dataset collects traffic traces of these $94$ apps from
    various network bandwidths (i.e., $30$ MB/s and $5$ MB/s) and network types
    (i.e., WiFi and cellular networks).
    The \underline{{\em SAMF-Failure}} dataset simulates network flapping caused by classical network failures, including
    packet loss, out-of-order, retransmissions, and flow loss, with various failure
    rates $0\sim30\%$.

\end{itemize}

\subsection{Single-App Fingerprinting Evaluation}
\label{subsec:single_app_evaluate}

\begin{table*}[t]
  \centering
  \small
  \caption{Precision(\%) and F1(\%) comparison on the FOAP dataset and MAPP dataset
      \label{tab:result_app_1}}
    \begin{tabular}{c|cc|cc|cc|cc|cc|cc|cc|cc}
    \bottomrule
    \multirow{3}{*}{Methods} &
    \multicolumn{8}{c|}{FOAP Dataset} &
    \multicolumn{8}{c}{MAPP Dataset}
    \\ \cline{2-17} &
    
    \multicolumn{2}{c|}{$w_t$=5s}  & \multicolumn{2}{c|}{$w_t$=10s}
    & \multicolumn{2}{c|}{$w_t$=20s}  & \multicolumn{2}{c|}{$w_t$=60s} &
    \multicolumn{2}{c|}{$w_t$=5s}  & \multicolumn{2}{c|}{$w_t$=10s}
    & \multicolumn{2}{c|}{$w_t$=20s}  & \multicolumn{2}{c}{$w_t$=60s}
    
    \\ \cline{2-17} &
    \multicolumn{1}{c|}{Pre} & \multicolumn{1}{c|}{F1} &
    \multicolumn{1}{c|}{Pre}  & \multicolumn{1}{c|}{F1} &
    \multicolumn{1}{c|}{Pre} & \multicolumn{1}{c|}{F1} &
    \multicolumn{1}{c|}{Pre}  & \multicolumn{1}{c|}{F1} &
    \multicolumn{1}{c|}{Pre}  & \multicolumn{1}{c|}{F1} &
    \multicolumn{1}{c|}{Pre} & \multicolumn{1}{c|}{F1} &
    \multicolumn{1}{c|}{Pre}  & \multicolumn{1}{c|}{F1} &
    \multicolumn{1}{c|}{Pre}  & \multicolumn{1}{c}{F1}
    \\
    \hline
AppScanner   & 77.8  & 72.1  & 81.2  & 74.8  & 86.3  & 80.5  & 92.4  & 87.1
             & 88.6  & 85.2  & 89.3  & 85.9  & 90.6  & 87.2  & 93.8  & 89.4\\
AWF          & 56.2  & 55.2  & 59.2  & 57.9  & 62.7  & 61.9  & 65.5  & 64.8
             & 68.5  & 68.0  & 71.0  & 70.7  & 72.1  & 71.8  & 75.1  & 75.0\\
DappGraph    & 61.3  & 59.8  & 62.8  & 60.9  & 63.4  & 61.4  & 66.8  & 64.0
             & 79.0  & 77.8  & 78.3  & 77.1  & 77.7  & 76.5  & 78.7  & 77.4\\
FS-Net       & 78.7  & 78.2  & 79.6  & 79.4  & 81.2  & 80.7  & 81.0  & 80.6
             & 80.1  & 79.3  & 78.1  & 77.3  & 80.7  & 79.0  & 78.7  & 78.2\\
FlowPrint    & 11.2  & 13.3  & 41.4  & 41.0  & 60.1  & 55.1  & 66.2  & 62.9
             & 67.2  & 36.2  & 67.7  & 45.8  & 69.4  & 55.3  & 81.0  & 75.0\\
MappGraph    & \underline{80.3}  & \underline{80.3}
             & \underline{85.2}  & \underline{85.1}
             & \underline{88.0}  & \underline{87.8}
             & \underline{90.7}  & \underline{90.6}
             & \underline{88.1}  & \underline{88.1}
             & 87.8  & 87.6
             & \underline{91.8}  & \underline{91.7}
             & 92.5  & 92.4\\
FG-Net       & 79.6  & 79.2  & 79.5  & 78.9  & 85.1  & 84.6  & 90.7  & 90.3
             & 87.8  & 87.8  & \underline{88.8}  & \underline{88.7}
             & 90.9  & 90.8  & \underline{93.6}  & \underline{93.3}\\
FlowMiner    & 76.1  & 74.5  & 80.9  & 80.5  & 84.3  & 83.8  & 90.4  & 90.4
             & 73.3  & 72.9  & 80.0  & 79.7  & 83.3  & 82.7  & 91.3  & 91.1\\  \hline
MAppPrint    & \textbf{85.1}  & \textbf{83.5}
             & \textbf{87.0}  & \textbf{86.1}
             & \textbf{91.5}  & \textbf{91.4}
             & \textbf{96.5}  & \textbf{96.4}
             & \textbf{92.3}  & \textbf{92.0}  & \textbf{93.7}  & \textbf{93.5}
             & \textbf{94.8}  & \textbf{94.8}  & \textbf{97.8}  & \textbf{97.8}\\
\toprule
\end{tabular}
\end{table*}

\begin{table*}[t]
  \centering
  \small
  \caption{Precision(\%) and F1(\%) comparison on the SAMF dataset and SIM-APP dataset
      \label{tab:result_app_2}}
    \begin{tabular}{c|cc|cc|cc|cc|cc|cc|cc|cc}
    \bottomrule
    \multirow{3}{*}{Methods} &
    \multicolumn{8}{c|}{SAMF Dataset} &
    \multicolumn{8}{c}{SIM-APP Dataset}
    \\ \cline{2-17} &
    
    \multicolumn{2}{c|}{$w_t$=5s}  & \multicolumn{2}{c|}{$w_t$=10s}
    & \multicolumn{2}{c|}{$w_t$=20s}  & \multicolumn{2}{c|}{$w_t$=60s} &
    \multicolumn{2}{c|}{$w_t$=5s}  & \multicolumn{2}{c|}{$w_t$=10s}
    & \multicolumn{2}{c|}{$w_t$=20s}  & \multicolumn{2}{c}{$w_t$=60s}
    
    \\ \cline{2-17} &
    \multicolumn{1}{c|}{Pre} & \multicolumn{1}{c|}{F1} &
    \multicolumn{1}{c|}{Pre}  & \multicolumn{1}{c|}{F1} &
    \multicolumn{1}{c|}{Pre} & \multicolumn{1}{c|}{F1} &
    \multicolumn{1}{c|}{Pre}  & \multicolumn{1}{c|}{F1} &
    \multicolumn{1}{c|}{Pre}  & \multicolumn{1}{c|}{F1} &
    \multicolumn{1}{c|}{Pre} & \multicolumn{1}{c|}{F1} &
    \multicolumn{1}{c|}{Pre}  & \multicolumn{1}{c|}{F1} &
    \multicolumn{1}{c|}{Pre}  & \multicolumn{1}{c}{F1}
    \\
    \hline
AppScanner   & 72.7  & 62.3  & 78.5  & 66.4  & 82.1  & 69.0  & 82.9  & 67.7

             & 77.1  & 68.8  & 78.3  & 67.9  & 78.9  & 65.4  & 84.9  & 69.4\\
             
AWF          & 73.6  & 72.4  & 75.2  & 73.7  & 75.0  & 74.3  & 76.7  & 75.6

             & \underline{70.5}  & \underline{70.1}
             & \underline{74.3}  & \underline{73.4}
             & \underline{75.3}  & \underline{74.6}
             & \underline{78.6}  & \underline{78.4}\\
DappGraph    & 50.7  & 48.2  & 52.4  & 47.4  & 51.9  & 46.5  & 50.7  & 44.6
             & 67.1  & 64.1  & 69.5  & 66.8  & 68.0  & 66.5  & 72.2  & 70.7\\
FS-Net       & 54.0  & 50.3  & 55.3  & 50.6  & 56.4  & 48.8  & 51.9  & 47.0
             & 68.9  & 66.6  & 69.4  & 67.9  & 70.3  & 68.7  & 72.3  & 70.8\\
FlowPrint    & 25.4  & 21.5  & 38.6  & 33.9  & 47.2  & 42.1  & 63.8  & 62.4
             & 18.6  & 18.2  & 24.6  & 24.4  & 46.3  &42.9  & 60.9  & 59.7\\
MappGraph    & \underline{76.5}  & \underline{75.8}
             & 76.9  & 76.7
             & 79.2  & 78.7
             & 81.1  & 80.8
             
             & 65.0  & 64.5
             & 67.3  & 67.0
             & 71.1  & 71.0
             & 76.3  & 76.3\\
             
FG-Net       & 74.6  & 74.3  & \underline{78.6}  & \underline{78.2}
             & 84.0  & 83.6  & \underline{88.0}  & \underline{87.7}
             
             & 70.6  & 69.6  & 71.1  & 70.3  & 71.4  & 71.4  & 77.3  & 77.2\\
FlowMiner    & 69.0  & 68.6  & 77.3  & 77.3  & \underline{85.0}  & \underline{84.6}
             & 86.4  & 86.2
             
             & 53.9  & 50.9  & 60.7  & 58.7  & 64.4  & 63.0  & 75.0  & 74.1\\  \hline
MAppPrint    & \textbf{80.2}  & \textbf{79.9}
             & \textbf{84.5}  & \textbf{84.3}
             & \textbf{89.0}  & \textbf{88.8}
             & \textbf{93.1}  & \textbf{93.1}
             
             & \textbf{77.4}  & \textbf{76.4}
             & \textbf{78.7}  & \textbf{77.1}
             & \textbf{79.7}  & \textbf{79.3}  & \textbf{87.1}  & \textbf{86.9}\\
\toprule
\end{tabular}
\vspace{-3ex}
\end{table*}

In the first experiment, we evaluate MAppPrint's performance on four single-app
datasets to answer RQ1.
For each dataset, we vary the traffic observation time window $w_t$ from $5$s to
$60$s, to investigate the trade-off between fingerprinting {\em accuracy} and
{\em timeliness}.

\bullethdr{App Identification in Normal Traces}.
\cref{tab:result_app_1} and
\cref{tab:result_app_2} present the results of all methods on three
{\em normal traffic} datasets, i.e., FOAP, MAPP, and SAMF.
MAppPrint consistently outperforms all baselines and achieves improvements of
$4.9\sim 5.8\%$ compared with the most competitive baselines, i.e., MappGraph
and FG-Net.
This superiority stems from MAppPrint's ability to capture unique in-app operational logic and module
dependencies, thus providing discriminative traffic fingerprints.
Meanwhile, the performance of each method generally improves and then reaches a plateau as $w_t$ increases.
The reason is that a longer observation time window helps collect richer traffic
fingerprinting information.
However, achieving enough accuracy with minimal $w_t$ is crucial for online app
fingerprinting, as $w_t$ directly determines fingerprinting latency
(see~\cref{subsec:efficiency_evaluate}).
We observe that, on these three datasets, MAppPrint's performance with $w_t=20$s
equals or surpasses all baselines when $w_t=60$s, indicating a timeliness
improvement over $300\%$.
This improvement is attributed to MAppPrint's ability to model traffic
fingerprints at a fine-grained traffic behavior level, enabling the acquisition
of rich app fingerprints even within a short observation time window.

We further investigate the causes of app misclassification.
In~\cref{fig:app_fpr} (see Appendix~\cref{sec:add_results}), we group the apps in the FOAP dataset into $12$ categories and observe that VPN, Video, Email, and News apps exhibit the highest false positive rates in fingerprinting. 
\Cref{tab:app_misclassification} (see Appendix~\cref{sec:add_results}) shows that misclassifications mainly occur between functionally similar apps or between apps from the same developer.
These misclassifications may arise because such apps use common CDNs and third-party SDKs, resulting in similar traffic patterns.

\bullethdr{App Identification in Ambiguous Traces}.
\Cref{tab:result_app_2} presents the results on a more challenging {\em ambiguous traffic} dataset SIM-APP.
MAppPrint achieves precision improvements of $6.9\sim 7.6\%$ and a six-fold
timeliness improvement, where MAppPrint's performance with $w_t=10$s equals or
exceeds that of all baselines with $w_t=60$s.

In particular, we observe that MAppPrint significantly outperforms existing multi-flow methods such as FG-Net and MappGraph.
The reason is that these methods construct fingerprints from flow-level concurrency patterns that may overlap across functionally similar apps, yielding ambiguous fingerprints.
In contrast, TBCG captures behavior-level traffic correlations that reflect an app’s module dependencies and operational logic, enabling MAppPrint to distinguish even functionally similar apps.

\begin{table}[htp]
    \centering
    \vspace{1ex}
    \small
    \caption{Precision(\%) and F1(\%) comparison in app-specific traffic and ambiguous traffic\label{tab:ambiguous}}
    \begin{tabular}{@{}c@{}c@{}|cc|cc|cc}
    \bottomrule
    \multicolumn{2}{c|}{\multirow{2}{*}{Methods}}  &
    \multicolumn{2}{c|}{FS-Net}  &
    \multicolumn{2}{c|}{MappGraph} &
    \multicolumn{2}{c}{MAppPrint} \\
    \cline{3-8} 
    \multicolumn{2}{c|}{}  &
    \multicolumn{1}{c|}{Pre} & \multicolumn{1}{c|}{F1} &
    \multicolumn{1}{c|}{Pre} & \multicolumn{1}{c|}{F1} &
    \multicolumn{1}{c|}{Pre} & \multicolumn{1}{c}{F1} \\ \hline
    
    \multicolumn{2}{c|}{App-specific traffic}  
    & \underline{84.0}  & \underline{83.5}  & 82.6  & 82.2  & \textbf{89.8}  & \textbf{89.6}  \\\hline  
\multicolumn{1}{@{}c@{}|}{\multirow{5}{*}{\rotatebox{-90}{\makecell{Ambiguous \\traffic }}}} 
& *.cloudfront.net   & \multicolumn{1}{c}{55.7}  & 49.6  & \multicolumn{1}{c}{\underline{62.0}}  & \underline{57.9}  & \multicolumn{1}{c}{\textbf{84.7}}  & \textbf{84.2}  \\ 
\multicolumn{1}{c|}{}                     
&  *.googleapis.com  & \multicolumn{1}{c}{\underline{41.6}}  & \underline{30.5}  & \multicolumn{1}{c}{38.5}  & 30.3  & \multicolumn{1}{c}{\textbf{79.4}}  & \textbf{75.9}  \\ 
\multicolumn{1}{c|}{}                     
& *.facebook.com    & \multicolumn{1}{c}{42.2}  &  29.7 & \multicolumn{1}{c}{\underline{45.8}}  & \underline{33.6}  & \multicolumn{1}{c}{\textbf{69.5}}  & \textbf{66.8}  \\  
\multicolumn{1}{c|}{}                     
& *.cloudflare.com  & \multicolumn{1}{c}{\underline{72.0}}  & \underline{68.3}  & \multicolumn{1}{c}{68.5}  & 61.9  & \multicolumn{1}{c}{\textbf{81.6}}  & \textbf{80.3}  \\ 
\multicolumn{1}{c|}{}                     
& *.doubleclick.net & \multicolumn{1}{c}{43.8}  & 36.2  & \multicolumn{1}{c}{\underline{45.1}}  & \underline{37.5}  & \multicolumn{1}{c}{\textbf{76.4}}  & \textbf{72.0}  \\ \hline
\toprule
\end{tabular}
\vspace{-3ex}
\end{table}

Furthermore, we divide the traffic traces in the SIM-APP dataset into {\em app-specific traffic}
and {\em cross-app shared ambiguous traffic} to quantify the impact of ambiguous traffic.
Such ambiguous traffic is typically generated by common CDNs,
third-party SDKs, and advertising services. 
\Cref{tab:ambiguous} compares the performance of different methods on app-specific traffic and top-$5$ ambiguous traffic.
We observe that the performance of existing methods significantly decreases in third-party SDKs (i.e., *.googleapis.com and *.facebook.com) and advertising (i.e., *.doubleclick.net) traffic, while in contrast, MAppPrint maintains optimal stability.
This demonstrates that, even for ambiguous traffic, TBCG 
can indeed capture app-specific dependencies, thereby supporting accurate fingerprinting.

\bullethdr{Ablation Analysis for TBCGs}.
We conduct an ablation analysis for two core components in TBCGs, i.e., unit traffic behaviors and behavior correlations, to evaluate their contributions to fingerprinting accuracy on the MAPP and SIM-APP datasets.
As shown in \cref{tab:tbcg-ablation} (see Appendix \cref{sec:add_results}), replacing unit traffic behaviors with conventional packet bursts~\cite{awf2018} or individually removing each type of behavior correlation reduces fingerprinting accuracy, with a particularly pronounced drop on the SIM-APP dataset.
These results demonstrate that both components are essential to the discriminability of TBCGs, particularly for apps with similar traffic patterns.

\begin{table*}[t]
  \centering
  \small
  \caption{Precision(\%) and F1(\%) comparison of foreground and background apps on the MAMF dataset
      \label{tab:result_app_3}}
    \begin{tabular}{c|cc|cc|cc|cc|cc|cc|cc|cc}
    \bottomrule
    \multirow{3}{*}{Methods} &
    \multicolumn{4}{c|}{$N_{a}=2$} &
    \multicolumn{4}{c|}{$N_{a}=3$} &
        \multicolumn{4}{c|}{$N_{a}=4$} &
    \multicolumn{4}{c}{$N_{a}=5$}
    \\ \cline{2-17} &
    
    \multicolumn{2}{c|}{Foreground}  & \multicolumn{2}{c|}{Background}
    & \multicolumn{2}{c|}{Foreground}  & \multicolumn{2}{c|}{Background} &
    \multicolumn{2}{c|}{Foreground}  & \multicolumn{2}{c|}{Background}
    & \multicolumn{2}{c|}{Foreground}  & \multicolumn{2}{c}{Background}
    
    \\ \cline{2-17} &
    \multicolumn{1}{c|}{Pre} & \multicolumn{1}{c|}{F1} &
    \multicolumn{1}{c|}{Pre}  & \multicolumn{1}{c|}{F1} &
    \multicolumn{1}{c|}{Pre} & \multicolumn{1}{c|}{F1} &
    \multicolumn{1}{c|}{Pre}  & \multicolumn{1}{c|}{F1} &
    \multicolumn{1}{c|}{Pre}  & \multicolumn{1}{c|}{F1} &
    \multicolumn{1}{c|}{Pre} & \multicolumn{1}{c|}{F1} &
    \multicolumn{1}{c|}{Pre}  & \multicolumn{1}{c|}{F1} &
    \multicolumn{1}{c|}{Pre}  & \multicolumn{1}{c}{F1}
    \\
    \hline
AppScanner   & 72.6  & 57.3  & 70.6  & 55.7  & 68.5  & 50.8  & 67.8  & 50.2  & 66.1  & 49.4  & 64.5  & 48.2  & 64.5  & 50.5  & 63.1  & 49.4\\
AWF          & \underline{68.2}  & \underline{61.8}  & \underline{66.9}  & \underline{60.6}  & \underline{67.9}  & \underline{59.1}  & \underline{66.9}  & \underline{58.2}  & \underline{66.2}  & \underline{56.3}  & \underline{64.8}  & \underline{55.1}  & \underline{66.3}  & \underline{56.0}  & \underline{64.7}  & \underline{54.7}\\
DappGraph    & 43.9  & 40.4  & 43.2  & 40.1  & 43.0  & 38.5  & 42.1  & 37.9  & 42.5  & 39.2  & 41.6  & 38.4  & 41.2  & 37.9  & 40.4  & 37.2\\
             
FS-Net       & 45.4  & 42.3  & 45.4  & 42.4  & 42.7  & 39.1  & 42.0  & 38.5  & 42.2  & 37.7  & 42.0  & 37.5  & 43.3  & 36.4  & 42.2  & 35.3\\
             
FlowPrint    & 37.0  & 32.9  & 28.8  & 25.1  & 34.2  & 29.4  & 25.9  & 21.6  & 26.4  & 22.0  & 19.9  & 16.5  & 20.6  & 14.6  & 15.8  & 11.4\\
MappGraph    & 44.5  & 37.7  & 37.5  & 31.7  & 40.8  & 35.3  & 34.4  & 29.8  & 39.7  & 30.4  & 34.8  & 26.7  & 27.7  & 19.1  & 23.8  & 16.7\\
FG-Net       & 55.2  & 47.6  & 47.8  & 41.2  & 51.1  & 45.8  & 43.7  & 39.1  & 41.4  & 35.1  & 35.6  & 30.2  & 34.1  & 26.8  & 29.0  & 22.8\\
FlowMiner     & 50.8  & 44.8  & 43.0  & 37.9  & 44.8  & 38.2  & 38.0  & 32.4  & 36.9  & 28.6  & 31.7  & 24.6  & 35.0  & 29.0  & 26.3  & 20.0   \\  \hline
MAppPrint    & \textbf{77.9}  & \textbf{75.7}  & \textbf{75.1}  & \textbf{73.0}  & \textbf{74.3}  & \textbf{70.6}  & \textbf{73.0}  & \textbf{69.4}  & \textbf{72.8 } & \textbf{69.9}  & \textbf{71.5}  & \textbf{68.6}  & \textbf{72.2}  & \textbf{68.1}  & \textbf{69.3}  & \textbf{65.4}\\
\toprule
\end{tabular}
\vspace{-2ex}
\end{table*}

\subsection{Multi-App Fingerprinting Evaluation}
\label{subsec:multi_app_evaluate}

In the next experiment, we answer RQ2 by evaluating MAppPrint on the multi-app dataset
MAMF, and we evaluate its reliability with varying numbers of concurrent
apps $N_{a}$.
The results in~\cref{tab:result_app_1,tab:result_app_2} show that a $30$s window provides a good trade-off between fingerprinting accuracy and timeliness. Therefore, we set $w_t=30$s in the following experiments. In practical deployments, $w_t$ can be adjusted according to traffic volume and latency requirements.

\bullethdr{App Identification in Multi-App Traces}.
In \cref{tab:result_app_3}, we compare the performance of different methods with $2\sim5$ concurrent apps. 
For each concurrency scenario, we evaluate the performance of {\em foreground} and {\em background} apps to analyze the impact of app running modes on fingerprinting.
Overall, we observe that MAppPrint is effective for multi-app fingerprinting, outperforming the best baseline by $12.1\sim 13.9\%$.
Comparing performance across different numbers of concurrent apps,
we can observe that as $N_a$ increases, the performance of multi-flow based baselines decreases significantly. 
This is because these methods are built on single-app assumptions and cannot distinguish mixed traffic from multiple concurrent apps, leading to incorrect modeling of traffic fingerprints.
In contrast, MAppPrint can separate the traffic of each individual app from mixed multi-app traces, thus demonstrating relatively stable performance across different $N_a$.

Furthermore, comparing foreground and background apps within the same $N_a$, we observe that foreground apps generally exhibit higher fingerprinting accuracy.
This may be because users can interact normally with foreground apps, resulting in traffic fingerprints that are similar to those in the single-app running mode. In contrast, users cannot directly interact with background apps; consequently, the resulting traffic traces reflect only partial fingerprint patterns.

\begin{figure*}[htbp]
	\centering
	\begin{minipage}[c]{0.33\textwidth}
		\centering
        \subfloat[Updated apps\label{fig:app_updata}]{            \includegraphics[width=0.5\linewidth]{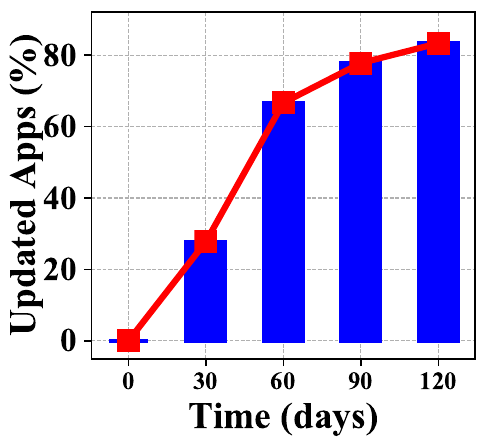}}
        \subfloat[Precision change\label{fig:pre_drift}]{            \includegraphics[width=0.5\linewidth]{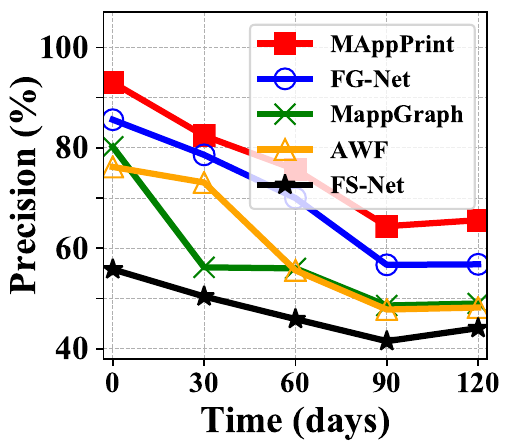}}
        \caption{Robustness in concept drift}
		\label{fig:concept_drift}
	\end{minipage} 
	\begin{minipage}[c]{0.66\textwidth}
		\centering
            \subfloat[Packet loss\label{fig:failure_pkt_loss}]{         \includegraphics[width=0.25\linewidth]{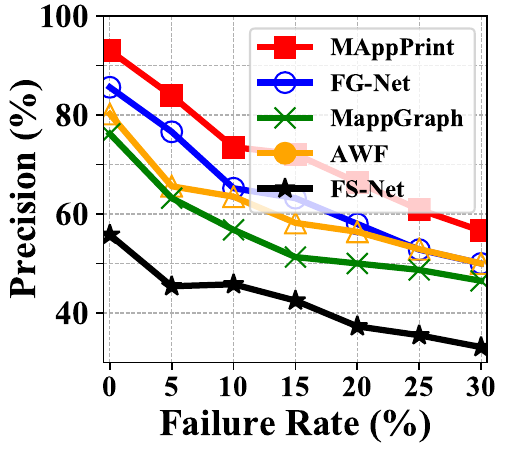}}
            \subfloat[Packet out-of-order\label{fig:failure_pkt_OOO}]{        \includegraphics[width=0.25\linewidth]{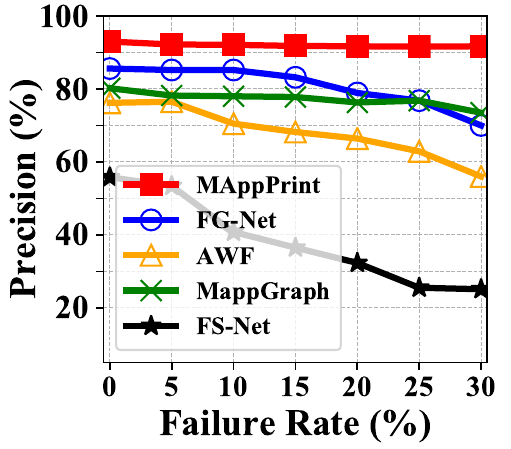}} 
            \subfloat[Packet retransmission\label{fig:failure_pkt_retrans}]{         \includegraphics[width=0.25\linewidth]{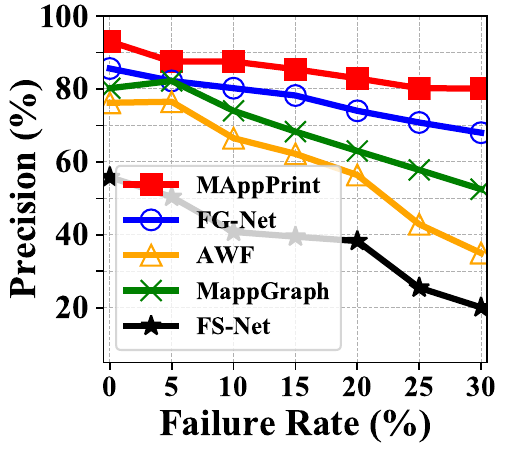}}
            \subfloat[Flow loss\label{fig:failure_flow_loss}]{        \includegraphics[width=0.25\linewidth]{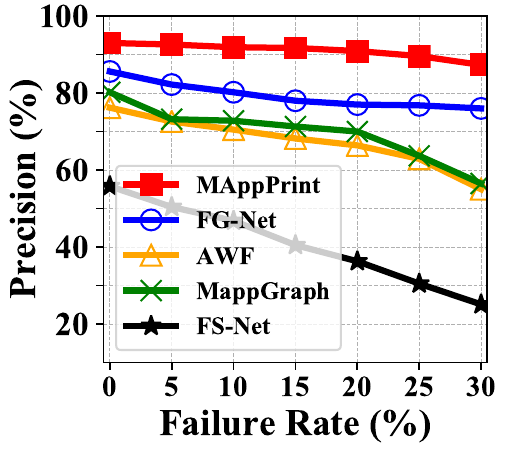}}
        \caption{Robustness in network flapping}
     \label{fig:failure_robust}
	\end{minipage}
    \vspace{-2ex}
\end{figure*}

In MAppPrint, multi-app fingerprinting relies on both the single-/multi-app traffic differentiation module and the multi-app traffic separation module.
In what follows, we further evaluate their effectiveness in app identification. 

\bullethdr{Effectiveness of Single-/Multi-App Traffic Differentiation}.
In \cref{tab:result_multi_identify}, we use the SAMF dataset and the MAMF dataset to evaluate whether MAppPrint can effectively distinguish single- and multi-app traffic traces.
We observe that MAppPrint achieves an overall single-/multi-app traffic differentiation precision of $92.1\sim 95.9\%$. 
We further group the traffic traces by traffic mixing ratio $c_r\in[0,0.5]$. 
As shown in \cref{tab:diff_mix} (see Appendix~\cref{sec:add_results}), 
MAppPrint achieves an overall precision of $90.1\sim 93.8\%$ in distinguishing single-app and multi-app traffic.
These results demonstrate that single-app traffic identification remains stable across different $N_a$ and $c_r$.
In contrast, the identification precision for multi-app traffic gradually improves as either $N_a$ or $c_r$ increases. 
This is because a more complex multi-app concurrency tends to more thoroughly
disrupt the traffic fingerprint of a multi-app traffic trace, making it
more distinct from single-app fingerprints.

\begin{table}[htp]
    \centering
    \small
    \caption{Performance of single-/multi-app differentiation \label{tab:result_multi_identify}}
\begin{tabular}{c|c|c@{ }|@{ }c@{ }|@{ }c@{ }|@{ }c@{ }}
\bottomrule
\multicolumn{2}{c|}{$\#app$}  
  & $N_{a} = 2$
  & $N_{a} = 3$
  & $N_{a} = 4$
  & $N_{a} = 5$
  \\ \hline
\multicolumn{1}{c|}{\multirow{2}{*}{\makecell{Single\\App}}}
                            & TPR(\%)  & 98.3 & 97.9  & 98.2 & 98.4 \\ \cline{2-2}
\multicolumn{1}{c|}{}       & FNR(\%)  & 1.7 &  2.1  & 1.8 & 1.6  \\ \hline
\multicolumn{1}{c|}{\multirow{2}{*}{\makecell{Multi-\\App}}}
                            & TPR(\%)  & 88.2 & 89.7  & 91.5 &  93.6  \\  \cline{2-2}
\multicolumn{1}{c|}{}       & FNR(\%)  & 11.8 & 10.3  & 8.5 &  6.4 \\  \hline
\multicolumn{1}{c|}{\multirow{2}{*}{Overall}}
                            & Pre(\%) & 92.1 & 93.4  & 95.7 &  95.9  \\   \cline{2-2}
\multicolumn{1}{c|}{}       & F1(\%)   & 91.0 & 92.6  & 93.3 &  93.5  \\
\toprule
\end{tabular}
\vspace{-1ex}
\end{table}

\bullethdr{Effectiveness of Multi-App Traffic Separation}.
In \cref{tab:result_ari}, we evaluate whether MAppPrint can effectively separate the traffic of different apps on multi-app traffic traces using the MAMF dataset.
We compare the clustering accuracy of different methods, where a higher value indicates a more effective separation of mixed multi-app traffic into single-app traffic. 
Since existing methods are defined for single-app scenarios, we extend them to multi-app traffic separation by combining their traffic fingerprints with the DBSCAN~\cite{schubert2017dbscan} algorithm.
MAppPrint achieves the best clustering accuracy across different $N_a$. 
This demonstrates that the multi-app traffic separation module can effectively decompose mixed multi-app traffic into individual single-app traffic.
As $N_a$ increases, the clustering accuracy exhibits a slight decline, which is consistent with our prior observation that a higher multi-app concurrency renders traffic patterns increasingly complex.

\begin{table}[htp]
  \centering
  \small
  \caption{Clustering accuracy (\%) of multi-app separation\label{tab:result_ari}}
  \begin{tabular}{c|c|c|c|c}
\bottomrule
    Methods & $N_a=2$ & $N_a=3$ & $N_a=4$ & $N_a=5$ \\
    \hline
    AWF   & 61.3    & 58.5       & 56.5      & 55.7        \\
    FS-Net   & 56.4    & 50.6       & 47.8      & 44.3    \\
    MappGraph   & 29.7    & 23.0       & 22.4      & 20.8    \\
    FG-Net   & 38.0    & 33.7       & 31.6      & 28.5    \\
    \hline
    MAppPrint    & 73.4    & 71.2       & 68.2      & 67.8         \\
    \toprule
  \end{tabular}
  \vspace{-2ex}
\end{table}

\subsection{Robustness Evaluation}
\label{subsec:robust_evaluate}

In the next experiment, we evaluate the robustness of different AF methods in {\em dynamic networking environments}.
We train each AF model in an initial environment, then we freeze
the parameters and evaluate the model in other environments without retraining.
We set the time window $w_t = 30$s.

\bullethdr{Concept Drift} refers to the phenomenon where data distribution
changes over time~\cite{han2023anomalydrift}.
In app fingerprinting, the {\em app updates} are the primary cause of concept drift.
We evaluate the performance of different methods when concept drift exists on
the SAMF-Drift dataset.
As shown in \cref{fig:app_updata}, the SAMF-Drift dataset tracked $94$ apps
continuously over a $120$-day period, where more than $80.0\%$ of the apps were
updated.
\cref{fig:pre_drift} demonstrates that MAppPrint is more robust than the other methods during this time period.
Even after $120$ days, it still exhibits a precision higher  than $65.0\%$.
This is because app updates typically involve only a few modules.
MAppPrint models app traffic generated via TBCG, which can capture
operational logic and module dependencies
in an app and thus can mitigate the impact of app updates.

\bullethdr{Regional Disparity and Network Changes}.
The performance of app fingerprinting is also affected by variations in model
{\em deployment regions}~\cite{prograph2023} and {\em network
bandwidth}~\cite{xie2023rosetta}.
We evaluate the performance of different methods with respect to these two
factors using the SAMF-Region and SAMF-Net datasets.
In~\cref{tab:robust_region}, ``A'', ``B'', and ``C'' are three
cities, where ``A'' is $50$ km from ``B'' and more than $1,000$ km from
``C''; ``H'', ``L'', and ``4G'' represent
$30$ MB/s wireless network, $5$ MB/s wireless network, and 4G cellular network.
The symbol ``$\Rightarrow$'' indicates the change in networking environment.
In general, when there are regional disparities or network changes, MAppPrint
performs much better than the other methods.

\begin{figure*}[htbp]
	\centering
	\begin{minipage}[c]{0.32\textwidth}
		\centering
        \subfloat[Throughput\label{fig:throughput}]{            \includegraphics[width=0.5\linewidth]{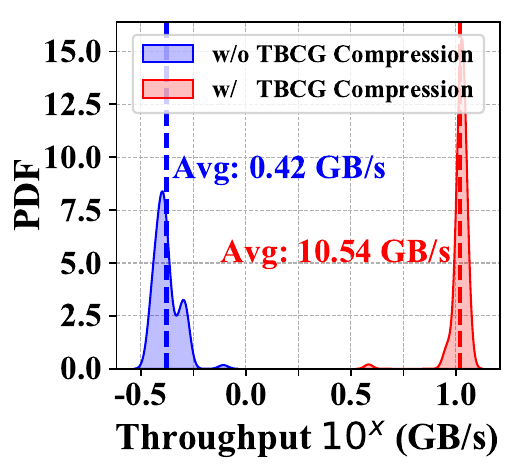}}
        \subfloat[Latency\label{fig:latency}]{            \includegraphics[width=0.5\linewidth]{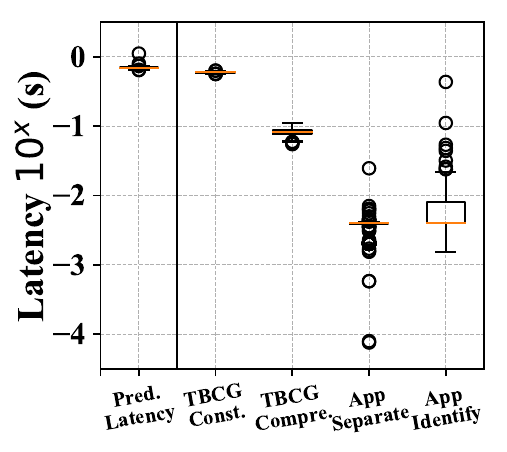}}
        \caption{Throughput and latency}
    \label{fig:throughput_latency}
	\end{minipage} 
	\begin{minipage}[c]{0.32\textwidth}
		\centering
        \subfloat[Graph scale reduction\label{fig:graph_scale}]{            \includegraphics[width=0.5\linewidth]{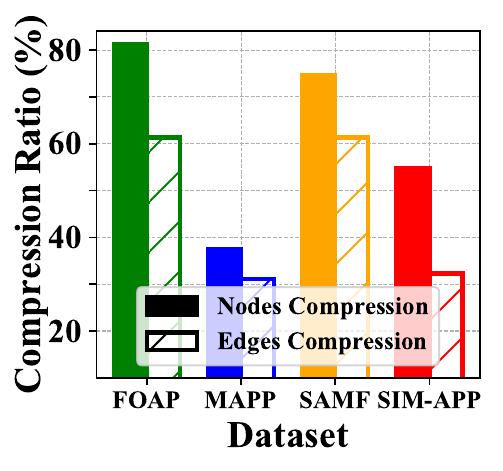}}
        \subfloat[Resource reduction\label{fig:resource}]{            \includegraphics[width=0.5\linewidth]{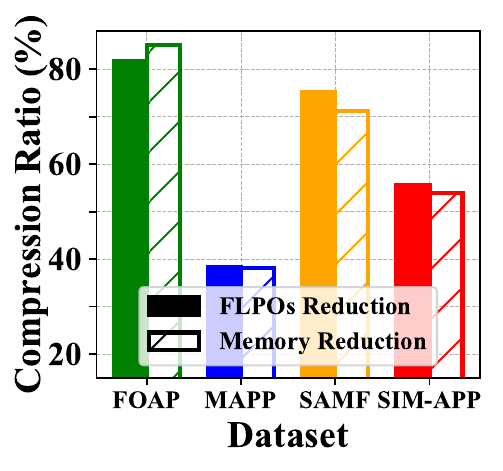}}
        \caption{Compression efficiencies}
    \label{fig:tbcg_compact}    
	\end{minipage}
	\begin{minipage}[c]{0.32\textwidth}
		\centering
        \subfloat[Precision\label{fig:pre_dense}]{            \includegraphics[width=0.5\linewidth]{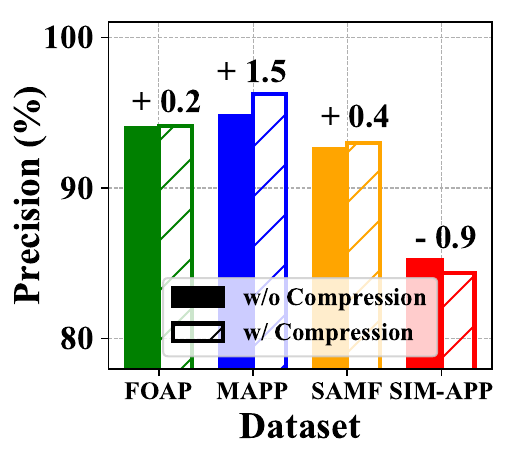}}
        \subfloat[Training efficiency\label{fig:train_dense}]{            \includegraphics[width=0.5\linewidth]{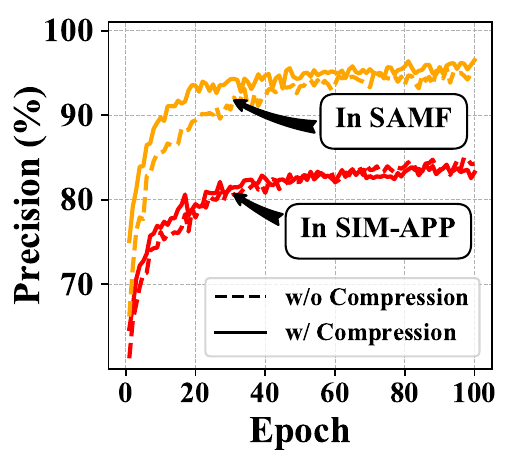}}
        \caption{Effects of compression}
    \label{fig:pre_compact}    
	\end{minipage}
\vspace{-2ex}
\end{figure*}

\begin{table}[htp]
  \centering
  \small
  \caption{Precision(\%) and F1(\%) in region/network changes}
  \label{tab:robust_region}
  \begin{tabular}{@{  }c@{  }|@{}c@{}c@{}|@{}c@{}c@{}|@{}c@{}c@{}|@{}c@{}c@{}}
    \bottomrule
    \multirow{3}{*}{Methods} &
    \multicolumn{4}{c|}{Regions Changes} &
    \multicolumn{4}{c}{Networks Changes}
    \\ \cline{2-9} 
    
    & \multicolumn{2}{c|}{A$\Rightarrow$B}
    & \multicolumn{2}{c|}{A$\Rightarrow$C}
    & \multicolumn{2}{c|}{H$\Rightarrow$L} 
    & \multicolumn{2}{c}{H$\Rightarrow$4G}\\
    \cline{2-9} 
    & \multicolumn{1}{c|}{Pre} & \multicolumn{1}{c|}{F1}
    & \multicolumn{1}{c|}{Pre} & \multicolumn{1}{c|}{F1}
    & \multicolumn{1}{c|}{Pre} & \multicolumn{1}{c|}{F1}
    & \multicolumn{1}{c|}{Pre} & \multicolumn{1}{c}{F1}\\
    \hline
    AWF         & 63.3  & 60.3  & 54.5 & 49.5  
                & 55.4  & 50.0  & 49.2   & 45.0\\
    FS-Net     & 56.7  & 47.0  & 43.7 & 33.2  
                &  47.6  & 44.8  & 45.4   & 43.6 \\
    MappGraph  & 65.9  & 61.3  & 47.2 & 43.1  
                &  46.6  & 46.1  & 42.9   & 39.8 \\
    FG-Net     
               & \underline{72.8} & \underline{69.2}
               & \underline{58.1} & \underline{52.0}  
               & \underline{64.3}  & \underline{58.6}
                & \underline{52.2}   & \underline{47.5}\\
    \hline
    MAppPrint  & \textbf{80.0}  & \textbf{76.3} 
               & \textbf{66.9}  & \textbf{60.2}
               & \textbf{72.3}  & \textbf{65.3}
              & \textbf{60.8}  & \textbf{53.6} \\
    \toprule
  \end{tabular}
\end{table}

\bullethdr{Network Flapping} caused by network failures such as packet loss,
packet out-of-order, packet retransmission, and flow loss can alter the traffic
patterns, posing substantial challenges to app fingerprinting.
We simulate such flapping in the SAMF-Failure dataset to evaluate the effect of {\em network flapping} on different AF methods.

The results shown in \cref{fig:failure_robust} indicate that
MAppPrint consistently demonstrates superior
robustness compared to other methods across various network failures,
exhibiting the least degradation in precision.
In particular, even under $30\%$ packet out-of-order or flow loss, the precision of MAppPrint degrades by merely $5\%$. This can be attributed to the fact that, although such network failures may alter the local feature patterns of network traffic, the correlations among traffic behaviors are associated with app module dependencies and thus preserve the discriminability of the fingerprints.

\bullethdr{Stability Analysis of Traffic Fingerprints}.
To understand the underlying reasons for MAppPrint's robustness, we conduct
a quantitative analysis of different traffic fingerprints' stability in
dynamic environments. In \cref{fig:cos_simi} (see Appendix \cref{sec:add_results}), we calculate
the average cosine similarity of traffic fingerprint representation vectors
for MAppPrint, MappGraph, and FS-Net in concept drift and region change
scenarios, where higher similarities indicate better fingerprint stability.
We observe that when the environment changes, the traffic representations of
all methods deviate from their initial states. Notably, MAppPrint's fingerprint
representation demonstrates significantly superior stability compared to other
methods. This suggests that the key to MAppPrint's robustness lies in the
proposed TBCG, which effectively models the invariant traffic patterns of apps
in dynamic environments.

\subsection{Efficiency Evaluation}
\label{subsec:efficiency_evaluate}

In the next experiment, we evaluate the efficiency of MAppPrint to answer RQ4.

\bullethdr{Throughput and Latency}.
We deploy MAppPrint on a laptop equipped with an i7-10875 CPU and RTX2060 GPU
to evaluate its throughput and latency on realistic computationally limited devices.
We replay the SAMF dataset to simulate {\em online app fingerprinting}.
As before, we set $w_t = 30$s.
The results are shown in \cref{fig:throughput_latency}.

\cref{fig:throughput} presents the {\em throughput} of MAppPrint, which achieves an
average throughput of $10.54$ GB/s.
This indicates that even in a high-speed wireless network of $30$ MB/s, MAppPrint
can simultaneously eavesdrop on approximately $360$ users' mobile devices using
only a laptop.
Notably, we also observe that the proposed TBCG compression module is a key factor
in achieving high throughput, significantly improving MAppPrint's throughput by
approximately $25$ times.

\cref{fig:latency} presents the {\em latency} of MAppPrint and its core modules.
MAppPrint presents an average {\em prediction latency} of $w_p=0.68$s, which
primarily arises from TBCG construction (average $0.60$s) and TBCG compression
(average $0.08$s) modules.
In contrast, multi-app traffic separation and app identification show only average latencies of $4.1\times10^{-3}$s and
$6.3\times10^{-3}$s.
In practical deployment, the {\em overall system latency} $w$ of an AF method
equals the sum of the size of traffic collection time window and the prediction
latency.
For MAppPrint, since $w_p \ll w_t$, its overall system latency $w = w_t + w_p \approx w_t$.
For real-time traffic traces, since $w_p \ll w_t$, MAppPrint has sufficient time
to process and identify traces from the previous window while collecting traces
for the current window.
This indicates that MAppPrint can effectively perform online AF.
In \cref{tab:all_throughput} (see Appendix~\cref{sec:add_results}), we further
compare the throughput and precision of different methods
at $30$s latency ($w_t = 30$s), where MAppPrint demonstrates at least a 65.2\% throughput
improvement and a 6.2\% precision improvement.

\bullethdr{Efficiency Analysis of TBCG compression}.
The above results indicate that the TBCG compression module is key to improving
MAppPrint's throughput.
We conduct a quantitative analysis of this module's {\em efficiency} by comparing the
graph scale of the traffic fingerprints and resources required for system
operation after and before TBCG compression.
The results shown in \cref{fig:tbcg_compact} indicate that TBCG compression can reduce the number of nodes by approximately $37.5\sim 81.5\%$ and the number of edges by $31.0 \sim 61.3\%$.
Meanwhile, it reduces floating-point operations (FLOPs) by $38.4 \sim 81.8\%$
and memory consumption by $38.1 \sim 85.1\%$ in the app identification step.

To evaluate the {\em impact of TBCG compression on fingerprinting accuracy},
we further compare the precision of MAppPrint with and
without the TBCG compression module.
The results shown in \cref{fig:pre_compact} indicate that an appropriate TBCG compression does not affect app fingerprinting precision very much.
Specifically, as shown in \cref{fig:pre_dense}, MAppPrint’s precision decreases by only $0.9\%$ on the SIM-APP dataset but increases by $0.2\%$, $1.5\%$, and $0.4\%$ on the other three datasets.
The reason is that TBCG compression only aggregates redundant nodes and edges,
and essential fingerprint information remains preserved.
The app identification model can thus focus more on the app-specific
information.
The effect of TBCG compression on the model training efficiency is presented in
\cref{fig:train_dense}, which is consistent with our previous observation.
During the rapid iteration phase (first $40$ epochs), MAppPrint trained on
compressed TBCGs consistently demonstrates higher precision than MAppPrint trained on uncompressed TBCGs.
We further evaluate the trade-off between compression ratio and fingerprinting accuracy on the MAPP and SAMF datasets. We adjust $K/|V|$ to obtain different compression ratios and evaluate the corresponding AF precision and system throughput. 
\Cref{fig:dynamic_compress} (see Appendix~\cref{sec:add_results}) shows that $K/|V|\in[0.1,0.2]$ provides a good balance, achieving compression ratios of $40\sim 80\%$ and increasing throughput by $22\sim25$ times, with only a $0.2\sim 0.9\%$ decrease in AF precision.
In contrast, lower compression ratios provide only limited throughput improvements, whereas excessive compression may substantially degrade AF precision.

\section{Discussion}
\label{sec:discussion}

In this section, we further discuss the generalizability of MAppPrint for other
application scenarios and explore possible countermeasures against MAppPrint.

\bullethdr{Fine-Grained In-app Activity Identification.}
In addition to identifying apps, fine-grained {\em in-app activity
inference}~\cite{saltaformaggio2016activity,zhang2023tfe}
is another important task in app fingerprinting.
We select $6$ high-interaction apps from the SAMF dataset 
to evaluate MAppPrint's performance in in-app activity identification. 
\Cref{tab:activity} (see Appendix \cref{sec:add_results}) compares the overall performance of MAppPrint with other methods,
with MAppPrint showing a lead of over $5\%$.
\Cref{tab:detail_activity} (see Appendix \cref{sec:add_results}) further presents the detailed effectiveness of MAppPrint for these activities.
We observe that MAppPrint can achieve an F1 score of
up to $98.5\%$ for certain in-app activities.

\bullethdr{Generalized Traffic Fingerprinting.}
More broadly, beyond app fingerprinting, {\em website
fingerprinting}~\cite{bah2023realWF}, {\em IoT
monitoring}~\cite{IotFP2020ma,li2025wifinger}, and {\em intrusion
detection}~\cite{fu2023hypervision,malicious2023fu} are also widely studied
problems in the field of traffic analysis.
We use three public datasets, i.e., X-Web~\cite{xiao2021ebsnn},
IoT-SENTINEL~\cite{xiao2021ebsnn}, and
USTC-TFC~\cite{wang2017malware}, to evaluate MAppPrint's generalizability to these
generalized traffic fingerprinting tasks.
As shown in \cref{tab:cross-domain} (see Appendix \cref{sec:add_results}),
MAppPrint demonstrates the best performance across different traffic fingerprinting tasks.
This indicates that MAppPrint is a universal traffic fingerprinting framework,
in which the proposed TBCG provides an effective cross-domain traffic fingerprint
modeling approach.

\bullethdr{Extend MAppPrint to Open World.}
In practical deployments, AF systems often face the {\em open world} challenge~\cite{li2022foap}.
This requires AF models to accurately distinguish between target and non-target app traffic,
rejecting non-target traffic instead of erroneously assigning it target app labels.
The multi-app traffic processing module enables MAppPrint to support open-world app fingerprinting.
Specifically, for an open-world traffic trace, we first use the single-/multi-app traffic differentiation module to identify whether it is a single-app trace generated by a target app.
For rejected traces, we further use the multi-app separation module to decompose them into multiple TBCG subgraphs and again determine whether each subgraph is a single-app trace generated by a target app.
If a subgraph is still rejected, it is considered to correspond to a non-target app; otherwise, it is regarded as traffic from a target app.
In \cref{tab:open} (see Appendix \cref{sec:add_results}), we evaluate MAppPrint's effectiveness in the open world. 
Results show that, despite a slight performance decrease, MAppPrint maintains its effectiveness in the open world.

\bullethdr{Impact of shared services.}
During app execution, services shared across apps (e.g., common CDNs and SDKs) may generate ambiguous traffic.
As discussed in~\cref{subsec:single_app_evaluate}, in single-app scenarios, the traffic dependencies modeled by MAppPrint provide app-specific context for ambiguous traffic, enabling MAppPrint to distinguish such traffic across apps.
In multi-app scenarios, however, ambiguous traffic may occasionally be misclustered during multi-app traffic separation. 
We observe that a limited degree of such misclustering does not necessarily lead to app misidentification because MAppPrint identifies apps based on reconstructed TBCGs rather than individual flows.
In future work, we plan to investigate MAppPrint's robustness under highly ambiguous traffic conditions.

\bullethdr{Impact of NAT.}
As with existing AF attacks~\cite{van2020flowprint,jiang2022fgnet,xu2025flowminer},
MAppPrint is designed to be deployed before network address translation (NAT), whereby the adversary can separate traffic flows generated by different devices in advance. In backbone networks, NAT and hierarchical routing can map flows from different devices within a subnet into flows with the same public IP but different ports. 
We simulate a NAT network using the SAMF dataset and analyze its impact.
The results in~\cref{tab:nat} (see Appendix~\cref{sec:add_results}) show that
when $2\sim 4$ devices behind a NAT are active simultaneously,
MAppPrint achieves a precision of $75.7\sim 81.6\%$ (a drop of $11.3\sim 17.3\%$).
We observe that MAppPrint identifies most multi-device traffic behind the NAT as multi-app traffic, separates the flows from different devices by clustering, and then identifies the apps. However, MAppPrint fails to correctly fingerprint NAT traffic when the same app runs simultaneously on different devices. We plan to investigate this issue in future work.

\bullethdr{Countermeasures against MAppPrint.}
MAppPrint poses a serious threat to user privacy.
It is thus important to explore efficient countermeasures.
Existing research has shown that traffic shaping techniques, including 
{\em packet padding}~\cite{luo2011httpos,wang2014effective}, 
{\em packet delaying}~\cite{bahramali2020practical,nasr2021defeating},
and {\em packet injection}~\cite{nasr2021defeating,gong2020zero},
are effective defenses against traffic fingerprinting.
In~\cref{fig:countermeasures} (see Appendix \cref{sec:add_results}), we analyze the
defensive capabilities of these three countermeasures against MAppPrint under
different defense intensities.
We observe that packet injection achieves the best defensive performance.
The reason is that packet injection not only alters the local patterns of
traffic behaviors but also disrupts their correlations.
Moreover, encrypted tunneling mechanisms (e.g., VPNs) provide another countermeasure against MAppPrint. This is because MAppPrint is built on traffic correlations, whereas these encrypted tunnels encapsulate multiple concurrent traffic flows within a single re-encrypted flow, making the correlations among the original traffic unobservable.
Meanwhile, we also notice that these countermeasures can result in substantial
bandwidth overhead and latency, potentially compromising the quality of service.

\section{Related Work}
\label{sec:related_work}

Traffic fingerprinting is an important branch of network traffic
analysis~\cite{chen2024mpaf}. It is widely applied in network
management~\cite{Qos2022wu,Qos2023cloud} and cybersecurity
maintenance~\cite{flowlens2021}, whereas it is also exploited to
eavesdrop on sensitive information~\cite{privacy2023}.

\header{App Fingerprinting (AF)}.
AF~\cite{wang2020appnet,oh2023appsniffer} aims to identify the apps activated by
users from encrypted traffic traces.
Existing AF methods can be categorized into {\em single-flow AF} and
{\em multi-flow AF}. Single-flow based methods model traffic fingerprints and
identify apps from independent traffic flows. For example,
AppScanner~\cite{taylor2016appscanner,taylor2017appscanner+}
represents each TCP flow using $54$ packet length statistical features and identifies
apps using an SVM classifier. DappGraph~\cite{shen2021dapp} and GFGC~\cite{gfgc2024}
represent each network flow as a packet interaction graph and identify apps
using a GNN. In contrast, multi-flow based methods model traffic fingerprints
from concurrent traffic flows generated by an app. For example,
FlowPrint~\cite{van2020flowprint} and MAppGraph~\cite{pham2021mappgraph} represent
concurrent flows as a destination correlation graph and identify apps through a graph
matching algorithm or a DGCNN~\cite{zhang2018dgcnn}.
FG-Net~\cite{jiang2022fgnet}, STC-WC~\cite{tan2024inter}, and FlowMiner~\cite{xu2025flowminer} represent concurrent
traffic flows as flow interaction graphs and identify apps using GNNs.

However, traditional single-flow AF methods fail to capture the correlations among traffic flows,
while current multi-flow AF methods suffer from vulnerability in dynamic environments, limited timeliness,
and an inability to handle multi-app scenarios.
Consequently, existing methods remain inefficient for realistic AF.

\header{Website Fingerprinting (WF)}.
WF~\cite{cherubin2022onlineWF,wang2022snwf} aims to identify the websites accessed
by users from encrypted traffic traces. Existing WF methods employ either manually designed
or automatically learned traffic fingerprints to achieve this goal. The former,
such as~\cite{hayes2016k-fp,panchenko2016cumul}, usually combine traffic feature
engineering with machine learning models like SVM and random forests to
design lightweight WF systems. The latter, such as~\cite{awf2018,bah2023realWF},
automatically learn fingerprint patterns and identify websites from raw traffic traces
by leveraging deep learning models like CNN, RNN, and Transformer.

Recently, multi-tab
WF~\cite{yin2021multi-tabWF,deng2023robustWF,jin2023multi-WF} has been proposed
to identify multiple websites accessed simultaneously by a user.
However, as discussed in \cref{ss:idea}, 
although these methods share a similar target with multi-app fingerprinting considered in our work, 
these methods model multi-website
identification as a multi-label classification problem and assume users can only
open a preset number of websites.
Such impractical assumptions impede their direct application in realistic AF.

\section{Conclusion}
\label{sec:conclusion}

In this work, we propose MAppPrint, a robust and real-time app fingerprinting method.
MAppPrint incorporates three key designs to fingerprint realistic mobile app traffic:
(i) a novel traffic modeling approach, i.e., TBCG, to model behavior-level
traffic correlations, providing
stable and
expressive app traffic patterns; (ii) an efficient TBCG compression algorithm to
improve throughput, enabling real-time app fingerprinting; and (iii) a unified app identification framework
to handle both single- and multi-app traffic traces.
Extensive experimental results demonstrate that MAppPrint achieves significantly
higher fingerprinting precision and timeliness than existing methods,
while exhibiting the highest robustness in dynamic networking environments and efficient real-time analysis in online deployment.

\section*{Ethical considerations}
\label{sec:ethical}

We meticulously adhered to ethical principles throughout our research process.
Considering that our research involves the analysis of sensitive network traffic,
we developed a traffic triggering system to {\em avoid} involving
{\em human subjects}. 
Current research has demonstrated that
app fingerprinting has both beneficial applications, such as improving app
security, and potentially harmful applications, like unauthorized eavesdropping. 
Hence, in our work, we thoroughly discussed possible countermeasures and made
efforts to minimize potential risks and harms.
In the collected datasets, we removed all sensitive content 
and anonymized the packet payloads. 
To mitigate potential misuse
of this technology, we have followed responsible disclosure practices, and all uses of our code and data will be subject to strict review and oversight.

\section*{Acknowledgements}
\label{sec:acknowledgements}

We thank our reviewers for their insightful comments and constructive feedback.
This work was supported by the National Natural Science Foundation of China (Grant No: 62272372), the Ministry of Science and Technology of the People's Republic of China (Grant No: 2025YFE0200100), and the Research Grants Council (Grant No: 15210023 and 15207725), Hong Kong SAR, China.

\bibliographystyle{IEEEtran}
\bibliography{ref}

\begin{appendices}

\section{Details of Implementation}
\label{sec:implementation}

\subsection{Unit Traffic Behavior Extraction}
\label{ss:behavior_extraction_alg}

\cref{alg:behavior_extraction} presents the procedure for extracting unit traffic behaviors, which segments a traffic trace $\CS$ into a set of unit 
traffic behaviors $\CB$.

\begin{algorithm}
  \caption{Unit Traffic Behavior Extraction\label{alg:behavior_extraction}}
  \KwIn{Traffic trace $\CS$, packet inter-arrival time threshold $\delta$ }
  \KwOut{A set of unit traffic behaviors $\CB$}
  $\CB\gets\emptyset$\;
  Group packets within $\CS$ into bidirectional flows $\{f_1,f_2,\dots\}$\;
  \tcp{Segment unit traffic behaviors within each flow}
  \ForEach{flow $f_i \in \{f_1,f_2,\dots\}$}{
    \ForEach{packet $pkt\in f_i$}{
      \uIf{$pkt$ is the first packet in flow $f_i$}{
        $b\gets\{pkt\}$\;
        $dir\gets pkt.dir$\;
        $time\gets pkt.time$\;
      }
      \uElseIf{$pkt.dir == dir$ and $pkt.time - time \leq \delta$}{
        $b\gets b\cup\{pkt\}$\;
        $time\gets pkt.time$\;
      }
      \Else{
        $\CB\gets\CB\cup\{b\}$\;
        $b\gets\{pkt\}$\;
        $dir\gets pkt.dir$\;
        $time\gets pkt.time$\;
      }
    }
    $\CB\gets\CB\cup\{b\}$\;
  }
\end{algorithm}

\subsection{Unit Traffic Behavior Features}
\label{ss:behavior_features}

\cref{tab:behavior_features} presents the features for vectorizing
unit traffic behaviors, which transforms each unit traffic behavior into a
$12$-dimensional feature vector.

\begin{table}[htp]
    \centering
    \small
    \caption{Unit traffic behavior features\label{tab:behavior_features}}
\begin{tabular}{c|c}
\bottomrule
      Attribute   & Description \\ \hline
\multirow{6}{*}{\makecell{Transmission\\ Features}}
                  &  Transmission protocol (TCP or UDP) \\
                  &  Direction (uplink or downlink)\\
                  &  Number of packets\\
                  &  Total number of bytes\\
                  &  Total duration\\
                  &  Transmission bit rate\\\hline
\multirow{3}{*}{\makecell{Packet Length\\ Features}}
                  &  Maximum packet length\\
                  &  Minimum packet length\\
                  &  Average packet length\\ \hline
\multirow{3}{*}{\makecell{Timing\\ Features}}
                  &  Maximum packet inter-arrival time\\
                  &  Minimum packet inter-arrival time\\
                  &  Average packet inter-arrival time\\

\toprule
\end{tabular}
  \vspace{5ex}
\end{table}

\subsection{TBCG Construction}
\label{ap:tbcg_construction}

\Cref{alg:tbcg_construction} presents the process of constructing a traffic behavior correlation graph, which transforms a unit traffic behavior set $\CB$ into a directed
acyclic graph $G$.

\begin{algorithm}
  \caption{Traffic Behavior Correlation Graph Construction\label{alg:tbcg_construction}}
  \KwIn{Unit traffic behavior set $\CB$, sequential correlation time threshold $\delta'$, triggering correlation time threshold $\delta''$ }
  \KwOut{A traffic behavior correlation graph $G = (V,E)$}
  Sort unit traffic behaviors in $\CB$ in ascending order according to their start times\;
  $V\gets\emptyset$, $E\gets\emptyset$\;
  \ForEach{unit traffic behavior $b_r^{(i)} \in \CB$}{
      $V\gets V\cup\{b_r^{(i)}\}$\;
      $start\_time = \text{Min}(b_r^{(i)}.time)$\tcp*[l]{Behavior start time}
      $end\_time = \text{Max}(b_r^{(i)}.time)$\tcp*[l]{Behavior end time}
      $max\_time = end\_time+\delta''$\tcp*[l]{Correlation range}
      $\CB_{corr} = \{b| b\in \CB, start\_time < \text{Min}(b.time) \leq max\_time \}$\;
      \ForEach{unit traffic behavior $b_s^{(j)} \in \CB_{corr}$}{
          $start\_time' = \text{Min}(b_s^{(j)}.time)$\;
          $end\_time' = \text{Max}(b_s^{(j)}.time)$\;
          \uIf{$i == j$ and $s == r+1$ and $start\_time' - end\_time \leq \delta'$}{
              $E\gets E\cup\{(b_r^{(i)}\rightarrow b_s^{(j)})\}$\tcp*[l]{Sequential edge}
          }
          \uElseIf{$i != j$ and $0\leq(start\_time' - end\_time)\leq \delta''$}{
              $E\gets E\cup\{(b_r^{(i)}\rightsquigarrow b_s^{(j)})\}$\tcp*[l]{Triggering edge}
          }
          \uElseIf{$i != j$ and $ start\_time' < end\_time $}{
               $E\gets E\cup\{(b_r^{(i)}\leftrightarrow  b_s^{(j)})\}$\tcp*[l]{Concurrent edge}
          }
          \Else{Continue\tcp*[l]{No edge exists}
          }
  }
}
\end{algorithm}

\subsection{Training of Single-App Classifiers}
\label{subsec:training}
\cref{alg:training} presents the training process of one-vs-rest single-app classifiers.

\subsection{Subgraph Features}
\label{subsec:subgraph}
\cref{tab:subgraph_features} shows the subgraph clustering features,
which transform each subgraph into a $10$-dimensional feature vector.

\begin{table}[htp]
    \centering
    \small
    \caption{Subgraph features\label{tab:subgraph_features}}
\begin{tabular}{c|c}
\bottomrule
   Attribute      & Description \\ \hline
\multirow{4}{*}{\makecell{Bidirectional\\ Features}}
                  &  Number of packets\\
                  &  Total bytes of packets\\
                  &  Total duration\\
                  &  Number of interactions\\ \hline
\multirow{6}{*}{\makecell{Unidirectional\\ Features}}
                  &  Number of uplink packets \\
                  &  Bytes of uplink packets\\
                  &  Average uplink packet length\\
                  &  Number of downlink packets\\
                  &  Bytes of downlink packets\\
                  &  Average downlink packet length\\

\toprule
\end{tabular}
    \vspace{5ex}
\end{table}

\begin{algorithm}
  \caption{Training of Single-app Classifiers\label{alg:training}}
  \KwIn{
        Training set $\CD = \{(G_1,y_1),(G_2,y_2),\dots \}$;
        model architecture $\Phi = \{\phi_1,\dots,\phi_L\}$;
        learning rate $l_r$
       }
  \KwOut{Model parameters $\Theta = \{\theta_1,\dots,\theta_L\}$}
  $\Theta \gets \emptyset$\;
  \tcp{Train each app classifier}
  \ForEach{classifier $\phi_i\in \Phi$}{
    \tcp{Construct binary classification dataset}
    $\CD_i\gets\emptyset$\;
    \ForEach{$(G_j,y_j)\in \CD$}{
    \uIf{$y_j == i$}{
        $\CD_i\gets \CD_i\cup\{(G_j,+1)\}$\tcp*[l]{Positive sample}
      }
      \Else{$\CD_i\gets \CD_i\cup\{(G_j,-1)\}$\tcp*[l]{Negative sample}}
    }
    Randomly initialize parameters of $\phi_i$ as $\theta_i$\;
    \tcp{Parameters updating}
    \ForEach{$(G_j,y_j)\in \mathcal{D}_i$}{
      $p_j = \phi_i(G_j|\theta_i)$\;
            $\theta_i \gets \theta_i - l_r\nabla \mathrm{CrossEntropy}(p_j,y_j)$\;
    }
    $\Theta \gets \Theta\cup\{\theta_i\}$\;
  }
\end{algorithm}

\section{Additional Experimental Setups}
\label{apdx:experiment_setup}

\begin{figure}[htp]
  \centering
  \includegraphics[width=0.8\linewidth]{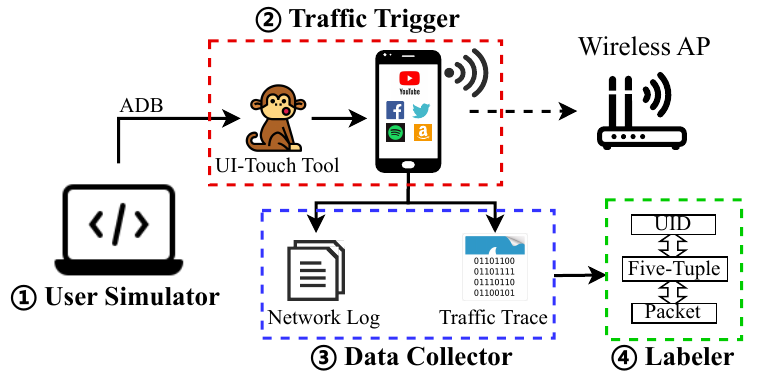}
  \caption{Workflow of traffic triggering and labeling system}
  \label{fig:traffic_collect}
  \end{figure} 

\subsection{Traffic Triggering and Labeling System}
\label{subsec:traffic_collect}

\Cref{fig:traffic_collect} presents the detailed framework and implementation
of our traffic triggering and labeling system.
The system comprises a user simulator,
a traffic trigger, a data collector, and a labeler.
The user simulator and labeler are implemented on a laptop,
while the traffic trigger and data collector are implemented on smartphones. 

The {\em user simulator} records and converts user activities into executable Android commands.
Various activity commands are combined, repeated to simulate continuous user
interactions, and sent to the traffic trigger through the Android Debug Bridge
(ADB).
The {\em traffic trigger} uses the UI-touch tool
Monkey~\cite{Monkey} to
automatically activate the corresponding apps and in-app components.
Simultaneously, the {\em data collector} records apps' system IDs (UIDs) and
five-tuples assigned to each app into network log files, and mirrors all
sent/received packets into PCAP files.
Finally, the {\em labeler} parses the invocation relationships between UIDs and
five-tuples recorded in the log files, and matches them with the packets'
five-tuples in the pcap files to achieve packet-level labeling.

\subsection{Model Architecture and Hyper-parameters Selection}
\label{sec:parameter}

We implement the app classifier using a GNN consisting of two stacked GAT layers, a graph pooling layer, and an output layer. \Cref{tab:gnn_architecture} presents the detailed model architecture.
We employ grid search to determine the optimal training hyper-parameters. 
\Cref{tab:parameter} presents the recommended hyper-parameter
configurations.

\begin{table}[htp]
    \centering
    \small
    \caption{Detailed GNN architecture}\label{tab:gnn_architecture}
\begin{tabular}{@{ }c@{ }|@{ }c@{ }|@{}c@{}|@{ }c@{ }}
\bottomrule
Module  & Network architecture  & Parameter    & Value \\ \hline
\multirow{4}{*}{\makecell{GAT \\Layers}}  
    & \multirow{4}{*}{\begin{tabular}[c]{@{}c@{}}\textcircled{\large 1} Multi-head GAT\\ 
    \textcircled{\large 2} ReLU \\ \textcircled{\large 3} Dropout \end{tabular}} 
       & Number of GAT layers  & 2     \\  &                             
       & Hidden dimension & 128   \\  &                         
       & Number of attention heads   & 4     \\ &                      
       & Dropout rate                    & 0.1   \\ 
\hline
\makecell{Graph\\Pooling \\Layer} & Mean pooling  & --   & --    \\ 
\hline
\multirow{2}{*}{\makecell{Output \\Layer}} 
    & \multirow{2}{*}{\makecell{\textcircled{\large 1} Single-layer MLP \\
    \textcircled{\large 2} Sigmoid}}       
    & Input dimension  & 128   \\   &  
       & Output dimension & 1     \\ 
\hline
\toprule
\end{tabular}
\vspace{3ex}
\end{table}

\begin{table}[htp]
    \centering
    \small
    \caption{Recommended hyper-parameter configurations}\label{tab:parameter}
\begin{tabular}{c|c|c}
\bottomrule
  \makecell{Hyper-\\parameter} & Description & Value \\ 
                \hline
                   $\delta$    & Packet inter-arrival time threshold          & $600$ms\\
                   $\delta'$   & Sequential correlation time threshold         & $1$s\\
                   $\delta''$   & Triggering correlation time threshold         & $2$s\\
                \hline
                   $K$         & Number of anchor behaviors      & $\#\text{apps}\times20$\\
                \hline
                   Eps$(\epsilon)$ & Maximum cluster radius                    & $0.5$\\
                   MinPts          & Minimum cluster points                    & $2$\\
                \hline
                   $l_r$       & Learning rate of model training               & $0.001$\\
                   BtSize      & Batch size of model training                  & $32$\\
\toprule
\end{tabular}
\vspace{3ex}
\end{table}

\section{Additional Experimental Results}
\label{sec:add_results}

\begin{figure}[htp]
  \centering
  \includegraphics[width=0.95\linewidth]{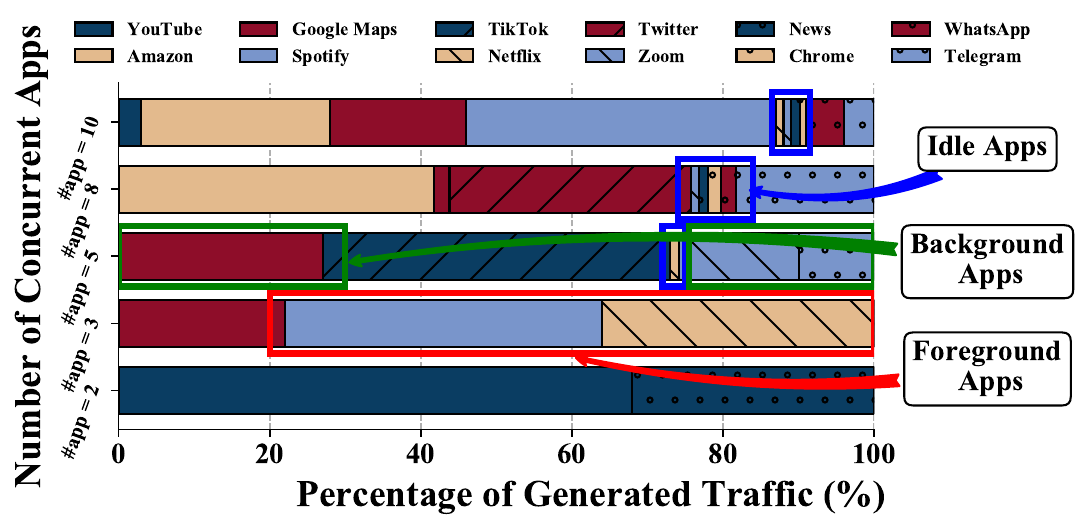}
  \caption{Distribution of traffic proportions under different multi-app concurrency scenarios}
  \label{fig:traffic_pattern}
\end{figure}

\begin{table}[htp]
    \centering
    \small
    \caption{Performance of single-/multi-app differentiation at different traffic mixing ratios $c_r$. The traffic mixing ratio denotes the fraction of packets from the least active app in the traffic collection window. \label{tab:diff_mix}}
\begin{tabular}{c|c|c@{ }|@{ }c@{ }|@{ }c@{ }|@{ }c@{ }|c}
\bottomrule
\multicolumn{2}{c|}{$c_r (\%)$}  
  & $\le 10$
  & $10\sim20$
  & $20\sim30$
  & $30\sim40$
  & $\geq 40$
  \\ \hline
\multicolumn{1}{c|}{\multirow{2}{*}{\makecell{Single\\App}}}
                            & TPR(\%)  & 97.6 & 97.5  & 97.8 & 97.6 & 97.6\\ \cline{2-2}
\multicolumn{1}{c|}{}       & FNR(\%)  & 2.4 &  2.5  & 2.2 & 2.4  & 2.4\\ \hline
\multicolumn{1}{c|}{\multirow{2}{*}{\makecell{Multi-\\App}}}
                            & TPR(\%)  & 86.8 & 87.9  & 90.6 &  91.3  & 92.2\\  \cline{2-2}
\multicolumn{1}{c|}{}       & FNR(\%)  & 13.2 & 12.1  & 9.4 &  8.7 & 7.8\\  \hline
\multicolumn{1}{c|}{\multirow{2}{*}{Overall}}
                            & Pre(\%) & 90.1 & 91.9  & 92.5 &  93.5  & 93.8\\   \cline{2-2}
\multicolumn{1}{c|}{}       & F1(\%)   & 89.3 & 90.6  & 91.8 &  92.3  & 93.0\\
\toprule
\end{tabular}
\end{table}

\begin{table}[htp]
    \centering
    \small
    \caption{Throughput (GB/s), Precision ($\%$), and F1 (\%) comparison of different methods at a $30$s
    latency\label{tab:all_throughput}}
\begin{tabular}{c|@{ }c@{ }|@{ }c@{ }|@{ }c@{ }|@{ }c@{ }|@{ }c@{ }}
\bottomrule
 Methods  & AWF & FS-Net & MappGraph & FG-Net & MAppPrint \\ 
\hline
 Throughput  & 2.87 & 0.90 & \underline{6.38} & 4.49 & \textbf{10.54} \\
 \hline
 Pre    & 76.4 & 54.2 & 80.9 & \underline{86.8} & \textbf{93.0} \\ 
 F1 & 75.9 & 47.3 & 80.0 & \underline{86.7} & \textbf{93.0} \\
\toprule
\end{tabular}
\end{table}

\begin{figure}[htp]
  \centering
                              \includegraphics[width=0.95\linewidth]{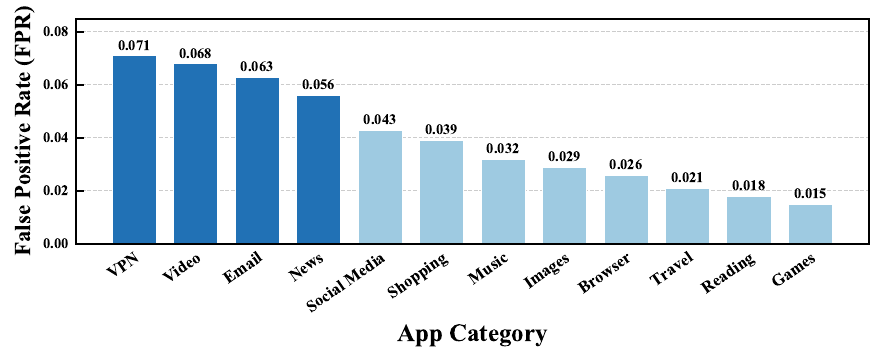}
  \caption{False positive rates by app categories on the FOAP dataset}
  \label{fig:app_fpr}
\end{figure}

\begin{table}[htp]
    \centering
    \small
                                        \caption{Incorrectly predicted labels and their proportions for the most frequently misclassified apps on the FOAP dataset\label{tab:app_misclassification}}
\begin{tabular}{@{}c@{}|@{}c@{}}
\bottomrule
      Apps  &  Incorrectly predicted labels (proportions)\\
      \hline
      kmvpn  & freenetvpn ($20.6\%$); freedomevpn ($5.9\%$); touchvpn ($4.5\%$)\\
      tubitv  & cardfeedvideo ($7.8\%$); mambettv ($5.3\%$); bigolive ($2.6\%$)\\
      yahoomail  & googlegm ($12.9\%$); saucefinder ($3.1\%$)\\
      tvbiNews & bingnews ($9.2\%$); bbcnews ($7.4\%$); nytimes ($5.0\%$)\\
      howtotie  & howtodraw ($14.6\%$); adobereader ($2.2\%$)\\

\toprule
\end{tabular}
\end{table}

\begin{figure}[htp]
  \centering
  \subfloat[{MAPP dataset}\label{fig:mapp_compress}]{    \includegraphics[width=0.5\linewidth]{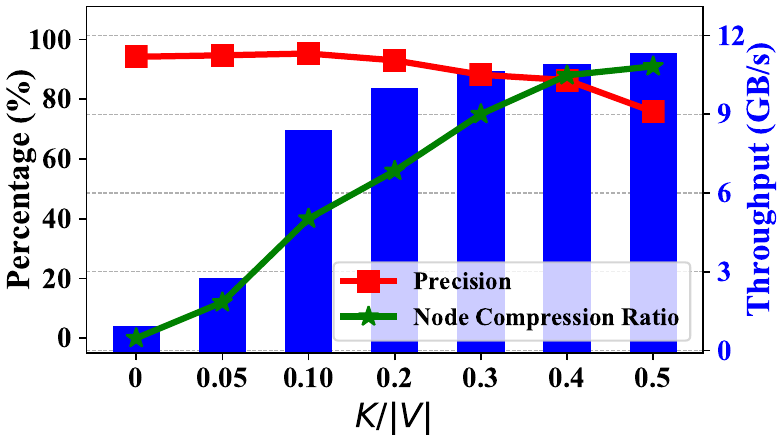}}
  \subfloat[SAMF dataset\label{fig:samp_compress}]{    \includegraphics[width=0.5\linewidth]{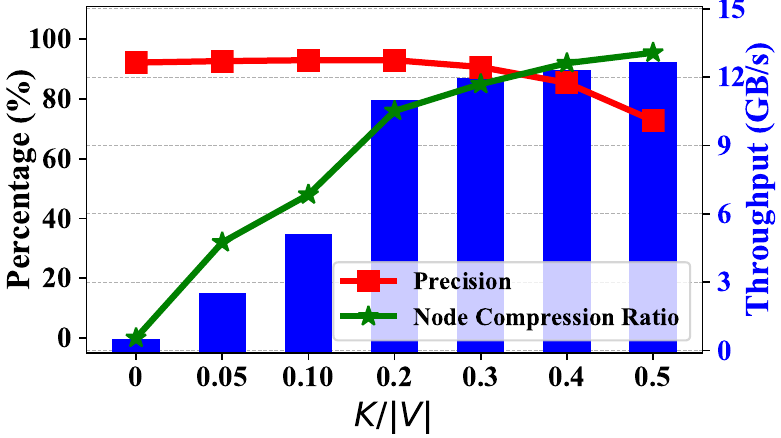}}
  \caption{Analysis of the dynamic trade-off between TBCG compression ratio and fingerprinting accuracy on the MAPP and SAMF datasets. Setting $K/|V|\in[0.1,0.2]$ provides a good balance.}
  \label{fig:dynamic_compress}
\end{figure}

\begin{figure}[t]
  \centering
  \subfloat[In concept drift\label{fig:simi_time}]{    \includegraphics[width=0.5\linewidth]{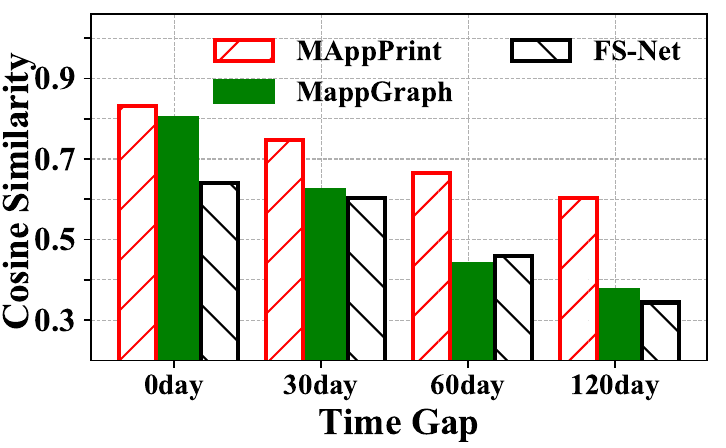}}
  \subfloat[In region change\label{fig:simi_region}]{    \includegraphics[width=0.5\linewidth]{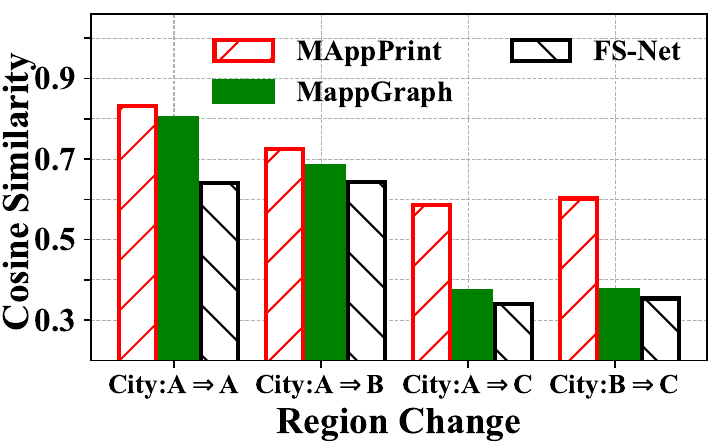}}
  \caption{Stability comparison of traffic fingerprints}
  \label{fig:cos_simi}
\end{figure}

\begin{table}[htp]
    \centering
    \small
    \caption{Performance comparison in the in-app activity identification task\label{tab:activity}}
\begin{tabular}{c|c|c|@{}c@{}|c|@{}c@{}}
\bottomrule
 Methods & AWF   & FS-Net & MappGraph & FG-Net             & MAppPrint \\ \hline
 Pre(\%)    & 43.6 & 48.1  &  62.3    &  \underline{67.4} &  \textbf{72.4}\\ 
 F1(\%)      & 41.0 & 39.6  &  61.5    &  \underline{67.3} &  \textbf{72.0} \\
\toprule
\end{tabular}
\end{table}

\begin{table}[htp]
\centering
\small
\caption{Detailed performance of MAppPrint for in-app activity identification
        \label{tab:detail_activity}}
\begin{tabular}{c|c|c|c}
\bottomrule
App                         & In-app Activity      &    Pre (\%)   & F1 (\%)     \\ \hline
\multirow{5}{*}{WhatsApp}   &  ChannelBrowse       &   50.5   &   49.2  \\  
                            &  TextChat            &   86.8   &  82.1   \\  
                            &  PictureSend         &   92.3   &   87.3  \\
                            &  VideoCall           &   92.9   &   83.9  \\ 
                            &  VoiceCall           &    98.3   & 96.7  \\ \hline
\multirow{5}{*}{\makecell{Telegram}}
                            &  ChannelBrowse       &   99.1   &   98.5    \\ 
                            &  TextChat            &   99.0   &   98.5    \\ 
                            &  PictureSend         &   93.9   &   92.5    \\ 
                            &  VideoCall           &   98.5   &   97.1  \\ 
                            &  VoiceCall           &   97.1   &   97.1   \\ \hline
\multirow{5}{*}{Shopee}     &  ProductSearch       &   89.7   &   81.3  \\ 
                            &  OrderBrowse         &   65.5   &   60.3   \\ 
                            &  ProductBrowse       &   58.6   &   58.6  \\                   
                            &  MessageSend         &   66.7   &   62.2   \\ 
                            &  ProductVideo        &   100.0  &   85.7  \\ \hline
\multirow{5}{*}{Amazon}     &  ProductSearch       &   83.3   &   87.0      \\ 
                            &  OrderBrowse         &   96.7   &   89.2   \\ 
                            &  ProductBrowse       &   89.6   &  83.3    \\ 
                            &  MessageSend         &   76.5   &  55.3   \\ 
                            &  ProductVideo        &   93.3   &  90.3   \\ \hline
\multirow{5}{*}{Twitter}    &  TweetBrowse         &   82.8   &  80.0   \\  
                            &  DirectMessages      &   100.0  &  88.4  \\ 
                            &  TopicsBrowse        &   87.1   &  87.1 \\ 
                            &  LiveView            &   46.5    &  41.7 \\ 
                            &  VideoBrowse         &   56.3   &  45.0 \\ \hline
\multirow{5}{*}{Instagram}  &  FeedBrowse          &   89.7   &   86.0  \\  
                            &  DirectMessages      &   92.5  &  90.8  \\ 
                            &  NoteBrowse          &   67.5   &  58.3 \\ 
                            &  LiveView            &   86.5   &  80.0 \\ 
                            &  VideoBrowse         &   91.8   &  84.6 \\ \hline
\textbf{Average}            &  \textbf{--}         &   72.4   &  72.0   \\ 
\toprule
\end{tabular}
\end{table}

\begin{table}[htp]
    \centering
    \small
    \caption{Precision (\%) and F1 (\%) in the website fingerprinting task (on the X-Web dataset), IoT monitoring task (on the IoT-SENTINEL dataset), and intrusion detection task (on the USTC-TFC dataset)\label{tab:cross-domain}}
    \begin{tabular}{c|cc|cc|cc}
    \bottomrule
    \multirow{2}{*}{Methods} &
    \multicolumn{2}{c|}{X-Web}  &
    \multicolumn{2}{c|}{IoT-SENTINEL} &
    \multicolumn{2}{c}{USTC-TFC} \\
    \cline{2-7} &
    \multicolumn{1}{c|}{Pre} & \multicolumn{1}{c|}{F1} &
    \multicolumn{1}{c|}{Pre} & \multicolumn{1}{c|}{F1} &
    \multicolumn{1}{c|}{Pre} & \multicolumn{1}{c}{F1} \\ \hline
AWF         & 60.9  & 60.7  & \underline{81.5}  & \underline{81.1}  & 95.7  & 95.2   \\  
FS-Net      & 69.4  & 65.0  & 75.4  & 74.4  & 77.0  & 75.7   \\ 
MappGraph   & 75.4  & 74.2  & 48.7  & 45.2  & 95.6  & 95.3   \\ 
FG-Net      & \underline{78.5}  & \underline{77.4}  & 74.2  & 72.6
            & \underline{99.5}  & \underline{99.4}\\  \hline
MAppPrint   & \textbf{80.0}  & \textbf{79.6}
            & \textbf{82.3}  & \textbf{81.6}
            & \textbf{99.5}  &\textbf{99.5}\\ 
\toprule
\end{tabular}
\end{table}

\begin{table}[htp]
    \centering
    \small
    \caption{Ablation analysis for TBCGs on the MAPP and SIM-APP datasets ($w_t = 30s$)\label{tab:tbcg-ablation}}
        \begin{tabular}{
        @{ }c@{ }|
        cc|
        cc
    }
    \bottomrule
        \multirow{2}{*}{Ablation}
        & \multicolumn{2}{c|}{MAPP}
        & \multicolumn{2}{c}{SIM-APP}\\
    \cline{2-5} &
    \multicolumn{1}{c|}{Pre} & \multicolumn{1}{c|}{F1} &
    \multicolumn{1}{c|}{Pre} & \multicolumn{1}{c}{F1} 
    \\ 
    \hline
\makecell{Replace unit traffic behaviors \\with packet bursts} & 94.7  & 94.5  & 82.4 &  81.8\\
\hline
Remove sequential edges         & 93.4  & 92.1  & 79.6  & 78.7     \\ 
Remove triggering edges         & 95.9  & 95.2  & 82.1  & 81.0     \\ 
Remove concurrent edges         & 94.4  & 94.0  & 81.3  & 80.2     \\ 
Remove all edges                & 92.8  & 91.5  & 77.8  & 75.4     \\

Use a fully connected graph      & 93.2  & 92.4  & 80.7  & 78.6     \\
\hline
  TBCGs                       & \textbf{97.2}  & \textbf{97.2}  & \textbf{84.8}  & \textbf{84.6}     \\

\toprule
\end{tabular}
\end{table}

\begin{table}[htp]
    \centering
    \small
    \caption{Precision (\%) and F1 (\%) comparison of MAppPrint in
    closed-world and open-world settings. 
    The closed-world dataset consists of $94$ target apps from the MAMF dataset.
    The open-world dataset consists of the closed-world dataset and $906$ non-target apps from the FOAP dataset.
    \label{tab:open}}
    \begin{tabular}{c|cc|cc|cc}
    \bottomrule
    \multirow{2}{*}{$N_a$} &
    \multicolumn{2}{c|}{$N_a=1$}  &
    \multicolumn{2}{c|}{$N_a=3$} &
    \multicolumn{2}{c}{$N_a=5$} \\
    \cline{2-7} &
    \multicolumn{1}{c|}{Pre} & \multicolumn{1}{c|}{F1} &
    \multicolumn{1}{c|}{Pre} & \multicolumn{1}{c|}{F1} &
    \multicolumn{1}{c|}{Pre} & \multicolumn{1}{c}{F1} \\ \hline
Closed world         & 93.0  & 93.0  & 73.6  & 70.0  & 70.6  & 66.8   \\  
Open world      & 91.9  & 90.3  & 68.6  & 64.9  & 66.3  & 62.5   \\ 
\toprule
\end{tabular}
\end{table}

\begin{figure}[htp]
  \centering
  \subfloat[Packet padding]{\includegraphics[width=0.32\linewidth]{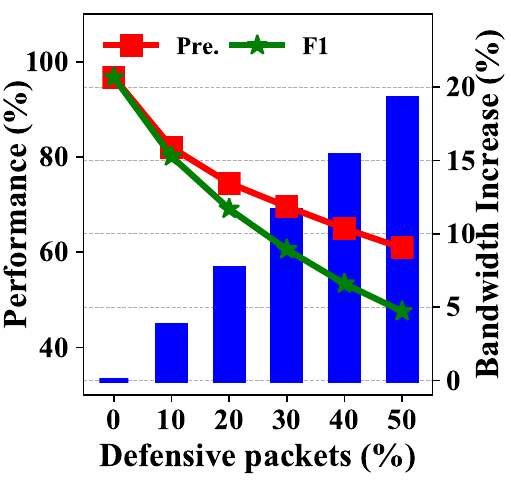}\label{fig:padding}}
  \subfloat[Packet delaying]{\includegraphics[width=0.32\linewidth]{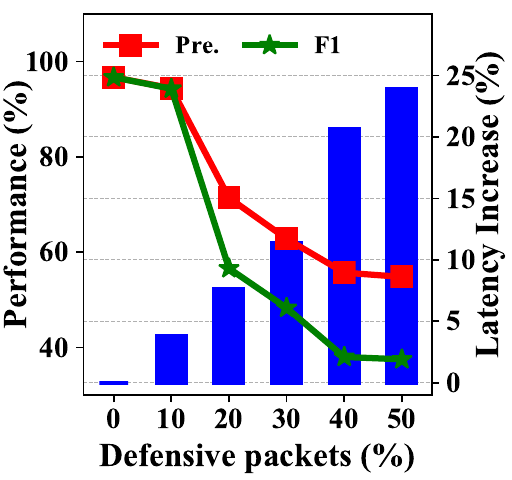}\label{fig:delaying}}
  \subfloat[Packet injection]{\includegraphics[width=0.32\linewidth]{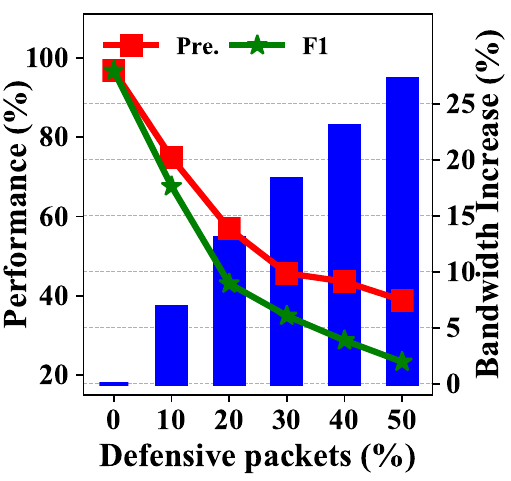}\label{fig:injecting}}
  \caption{Performance of MAppPrint in countermeasures
    \label{fig:countermeasures}}
\end{figure}

\begin{table}[htp]
    \centering
    \small
    \caption{Precision (\%) and F1 (\%) of MAppPrint in a
    simulated NAT network. The IP addresses of different devices are mapped to the same public IP address, distinct ports are assigned to each device, and the start timestamps of their traffic traces are aligned.
    $N_d$ denotes the number of devices behind the NAT; ``w/o NAT'' (``w/ NAT'') denotes the setting without (with) NAT.\label{tab:nat}}
    \begin{tabular}{c|cc|cc|cc}
    \bottomrule
    \multirow{2}{*}{$N_d$} &
    \multicolumn{2}{c|}{$N_d=1$}  &
    \multicolumn{2}{c|}{$N_d=2$} &
    \multicolumn{2}{c}{$N_d=4$} \\
    \cline{2-7} &
    \multicolumn{1}{c|}{Pre} & \multicolumn{1}{c|}{F1} &
    \multicolumn{1}{c|}{Pre} & \multicolumn{1}{c|}{F1} &
    \multicolumn{1}{c|}{Pre} & \multicolumn{1}{c}{F1} \\ \hline
w/o  NAT         & 93.0  & 92.9  & 92.9  & 92.8  & 93.0  & 93.0   \\  
w/   NAT     & 93.0  & 93.0  & 81.6  & 80.5  & 75.7  & 73.4   \\ 
\toprule
\end{tabular}
\end{table}
\end{appendices}

\end{document}